\documentclass{article}

\usepackage{PRIMEarxiv}

\usepackage[utf8]{inputenc} 
\usepackage[T1]{fontenc}    
\usepackage[hidelinks]{hyperref}  
\usepackage{hyperref}       
\usepackage{url}            
\usepackage{booktabs}       
\usepackage{amsfonts}       
\usepackage{nicefrac}       
\usepackage{microtype}      
\usepackage{lipsum}
\usepackage{fancyhdr}       
\usepackage{graphicx}       
\graphicspath{{media/}}     

\usepackage{pdflscape}
\usepackage{afterpage}
\usepackage{colortbl}
\usepackage{makecell}
\usepackage{multirow}
\usepackage{subcaption}

\newcommand\blfootnote[1]{%
  \begingroup
  \renewcommand\thefootnote{}\footnote{#1}%
  \addtocounter{footnote}{-1}%
  \endgroup
}

\newcommand{\changed}[1]{#1}
  
\title{AIROGS: Artificial Intelligence for RObust Glaucoma Screening Challenge
}

\author{Coen~de~Vente, Koenraad~A.~Vermeer, Nicolas~Jaccard, He~Wang, Hongyi~Sun, Firas~Khader, Daniel~Truhn, \And Temirgali~Aimyshev, Yerkebulan~Zhanibekuly, Tien-Dung~Le, Adrian~Galdran, \And Miguel~\'{A}ngel~Gonz\'{a}lez Ballester,  Gustavo~Carneiro, Devika~R~G, Hrishikesh~P~S, Densen~Puthussery, \And Hong~Liu, Zekang~Yang, Satoshi~Kondo, Satoshi~Kasai, Edward~Wang, Ashritha~Durvasula, \And J\'{o}nathan~Heras, Miguel~\'{A}ngel~Zapata, Teresa~Ara\'{u}jo, Guilherme~Aresta, Hrvoje~Bogunovi\'{c}, \And Mustafa~Arikan, Yeong~Chan~Lee, Hyun~Bin~Cho, Yoon~Ho~Choi, Abdul~Qayyum, Imran~Razzak, \And Bram~van~Ginneken, Hans~G.~Lemij, Clara~I.~S\'{a}nchez}

\begin{document}
\maketitle

\begin{abstract}
The early detection of glaucoma is essential in preventing visual impairment. Artificial intelligence (AI) can be used to analyze color fundus photographs (CFPs) in a cost-effective manner, making glaucoma screening more accessible. While AI models for glaucoma screening from CFPs have shown promising results in laboratory settings, their performance decreases significantly in real-world scenarios due to the presence of out-of-distribution and low-quality images. To address this issue, we propose the Artificial Intelligence for Robust Glaucoma Screening (AIROGS) challenge. This challenge includes a large dataset of around 113,000 images from about 60,000 patients and 500 different screening centers, and encourages the development of algorithms that are robust to ungradable and unexpected input data. We evaluated solutions from 14 teams in this paper, and found that the best teams performed similarly to a set of 20 expert ophthalmologists and optometrists. The highest-scoring team achieved an area under the receiver operating characteristic curve of 0.99 (95\% CI: 0.98-0.99) for detecting ungradable images on-the-fly. Additionally, many of the algorithms showed robust performance when tested on three other publicly available datasets. These results demonstrate the feasibility of robust AI-enabled glaucoma screening.

\end{abstract}


\blfootnote{This research was
funded in part by Eurostars grant E12712 and supported in part by
Amazon Web Services. (Corresponding author: Coen de Vente.)
Please see the Acknowledgment section of this article for the author
affiliations.} 

\section{Introduction}
\label{sec:introduction}
Glaucoma is one of the main causes of irreversible blindness and impaired vision in the world. It affects the optic nerve, which connects the eye with the brain, and leads to progressive visual field damage. This damage initially passes unnoticed by the patient. Only in later stages will glaucoma patients experience visual loss. According to estimates, by 2040, over 110 million people will have varying degrees of visual impairment caused by glaucoma \cite{Tham14}, with 10\% becoming blind in both eyes and 25\% in one eye \cite{Mokh16}. Many people experience visual impairment from glaucoma because it is often not detected until later stages \cite{Erne13,Pete13}. Current treatments of glaucoma cannot repair the damage, but can only halt or slow the progression of the condition \cite{Wein14}. Implementing screening programs to identify patients early on for treatment can alleviate the consequences of the disease. Artificial intelligence (AI) may be the enabling technology for the cost-effective implementation of these programs by automatically detecting perimetric glaucoma (i.e., glaucoma in which there is already visual field damage) in color fundus photographs (CFPs) \cite{Chen15,Ting17,Li18,Phen19,Roge19}.

\begin{table*}[!t]
\centering
\caption{Statistics of the Rotterdam EyePACS AIROGS dataset. \# = number of. CFPs = color fundus photographs.}

\resizebox{\textwidth}{!}{\begin{tabular}{@{}rrrrrr@{}}

\toprule
\multicolumn{1}{l}{}      & \multicolumn{1}{l}{} & \multicolumn{1}{l}{} & Full set                  & Training set                    & Test set                   \\

\midrule
\midrule
 
\multicolumn{3}{r}{\# CFPs / \# patients}                               & 112,732 / 60,071          & 101,442 / 54,274         & 11,290 / 5,797         \\

\midrule
\multicolumn{2}{r}{\multirow{3}{*}{\begin{tabular}[c]{@{}r@{}}Prevalence\\ (\# CFPs (\% within set) / \# patients)\end{tabular}}}  & RG                   & 4,872 (4.3\%) / 3,531     & 3,270 (3.2\%) / 2,336    & 1,602 (14.2\%) / 1,195 \\
\multicolumn{2}{r}{}                             & NRG                  & 106,306 (94.3\%) / 58,388 & 98,172 (96.8\%) / 53,400 & 8,134 (72.0\%) / 4,988 \\
\multicolumn{2}{r}{}                             & U                    & 1,554 (1.4\%) / 1,415     & 0 (0.0\%) / 0            & 1,554 (13.8\%) / 1,415 \\

\midrule
\multicolumn{3}{r}{Age at encounter (mean $\pm$ std. dev.)}             & 56.9 $\pm$ 10.3
                          & 56.7 $\pm$ 10.2          & 59.3 $\pm$ 10.9        \\

\midrule
\multicolumn{3}{r}{\# sites}                                            & 500                       & 486                      & 376                    \\

\midrule
\multirow{10}{*}{\multirow{2}{*}{\begin{tabular}[c]{@{}r@{}}Cameras\\ (\# CFPs (\% within set) / \# patients)\end{tabular}}} & Canon                & CR1                  & 11,462 (10.2\%) / 6,013   & 10,274 (10.1\%) / 5,422  & 1,188 (10.5\%) / 591   \\
                          & Canon                     & CR2                  & 10,523 (9.3\%) / 5,538    & 9,179 (9.0\%) / 4,866    & 1,344 (11.9\%) / 672   \\
                          & Canon                     & DGI                  & 10,644 (9.4\%) / 5,690    & 9,581 (9.4\%) / 5,145    & 1,063 (9.4\%) / 545    \\
                          & Optovue              & iCam 100             & 29,108 (25.8\%) / 16,166  & 26,480 (26.1\%) / 14,742 & 2,628 (23.3\%) / 1,424 \\
                          & TopCon               & NW200                & 3,109 (2.8\%) / 1,588     & 2,888 (2.8\%) / 1,478    & 221 (2.0\%) / 110      \\
                          & TopCon                     & NW400                & 22,519 (20.0\%) / 11,736  & 20,557 (20.3\%) / 10,737 & 1,962 (17.4\%) / 999   \\
                          & Centervue            & DRS                  & 1,805 (1.6\%) / 988       & 1,598 (1.6\%) / 879      & 207 (1.8\%) / 109      \\
                          & Nidek                & AFC300               & 61 (0.1\%) / 31           & 53 (0.1\%) / 27          & 8 (0.1\%) / 4          \\
                          & Crystalvue           & NFC-700              & 8 (0.0\%) / 4             & 6 (0.0\%) / 3            & 2 (0.0\%) / 1          \\
                          & Unknown              &                      & 23,493 (20.8\%) / 12,323  & 20,826 (20.5\%) / 10,981 & 2,667 (23.6\%) / 1,342 \\
                          \bottomrule
\end{tabular}}
\label{table:dataset_stats}
\end{table*}

Existing AI solutions have been shown to drop in performance in real-world screening practice due to comorbidities, poor quality images, different ethnicities, or unexpected out-of-distribution (OOD) samples \cite{Phen19}. Ad-hoc quality check modules have been added on AI solutions to overcome this performance drop, but recent research has indicated that these quality checks are not sufficiently accurate when deployed in real-world settings \cite{Beed20}.
To allow a safe and effective deployment in screening, the reliability and robustness of such solutions need to be assessed. 
Medical image analysis challenges often exclusively focuses on performance metrics that are potentially unrealistic and overestimated due to the use of test sets that do not represent real-world scenarios. Moreover, metrics to measure reliability and robustness are often neglected due to the difficulty to estimate them in the provided test sets. 

With the aim to develop solutions that overcome the aforementioned issues related to robustness in glaucoma screening, we organized the Artificial Intelligence for RObust
Glaucoma Screening (AIROGS) challenge. The goal of this challenge was to evaluate the feasibility of the development of a state-of-the-art, reliable AI solution that takes a CFP as input and provides as output the likelihood of referable glaucoma, accompanied with outputs for robustness (i.e., predicting whether the input image can be graded reliably or not).

To encourage the development of solutions that are robust to any kind of ungradable and unexpected input data and are equipped with inherent robustness mechanisms, the training set we provided was a subset of the full AIROGS dataset where only gradable images were included and ungradable images excluded. The test set, however, is unfiltered, containing all images found in screening settings (gradable and ungradable), representing a real-world scenario.

AIROGS was part of the International Symposium on Biomedical Imaging (ISBI) 2022 challenge program. It reopened after presenting the results during ISBI 2022 and submissions can still be made on Grand Challenge\footnote{https://airogs.grand-challenge.org}.

Our challenge, along with the dataset we made publicly available, distinguishes itself from previous glaucoma challenges.
First, to the best of our knowledge, our dataset is the largest publicly available CFP dataset with glaucoma labels by a large margin. In total, our dataset contains 112,732 CFPs, exceeding the size of other publicly available datasets containing CFPs with glaucoma, of which the sizes range from 22 to 2,000 CFPs \cite{Zhan10,Fume11,Odst13,Siva15,Holm17,Orla18,Orla20,Fang22,Wu22}. It is a highly diverse dataset as it originates from 500 screening centers across the United States of America and was acquired with a large variety of cameras.
Second, the AIROGS challenge is the first challenge to emphasize robustness in glaucoma screening.
Third, AIROGS is one of the first type of challenges on grand-challenge.org that requires participants to submit an algorithm (a \textit{Type 2} challenge), rather than a file with their predictions on the test set (a \textit{Type 1} challenge), as is done in more traditional challenges. This makes human intervention in the generation of test set results impossible, reducing the possibility of cheating. Morever, it greatly improves reproducibility, allowing everyone to reuse the trained algorithms that were submitted and apply them to new data in a cloud-based environment.
Fourth, the reproducibility enabled testing of the participating algorithms on three external datasets: two for evaluating the screening task and one for evaluating robustness.

\section{Datasets}
\begin{figure*}[!t]
\centerline{\includegraphics[width=.95\linewidth]{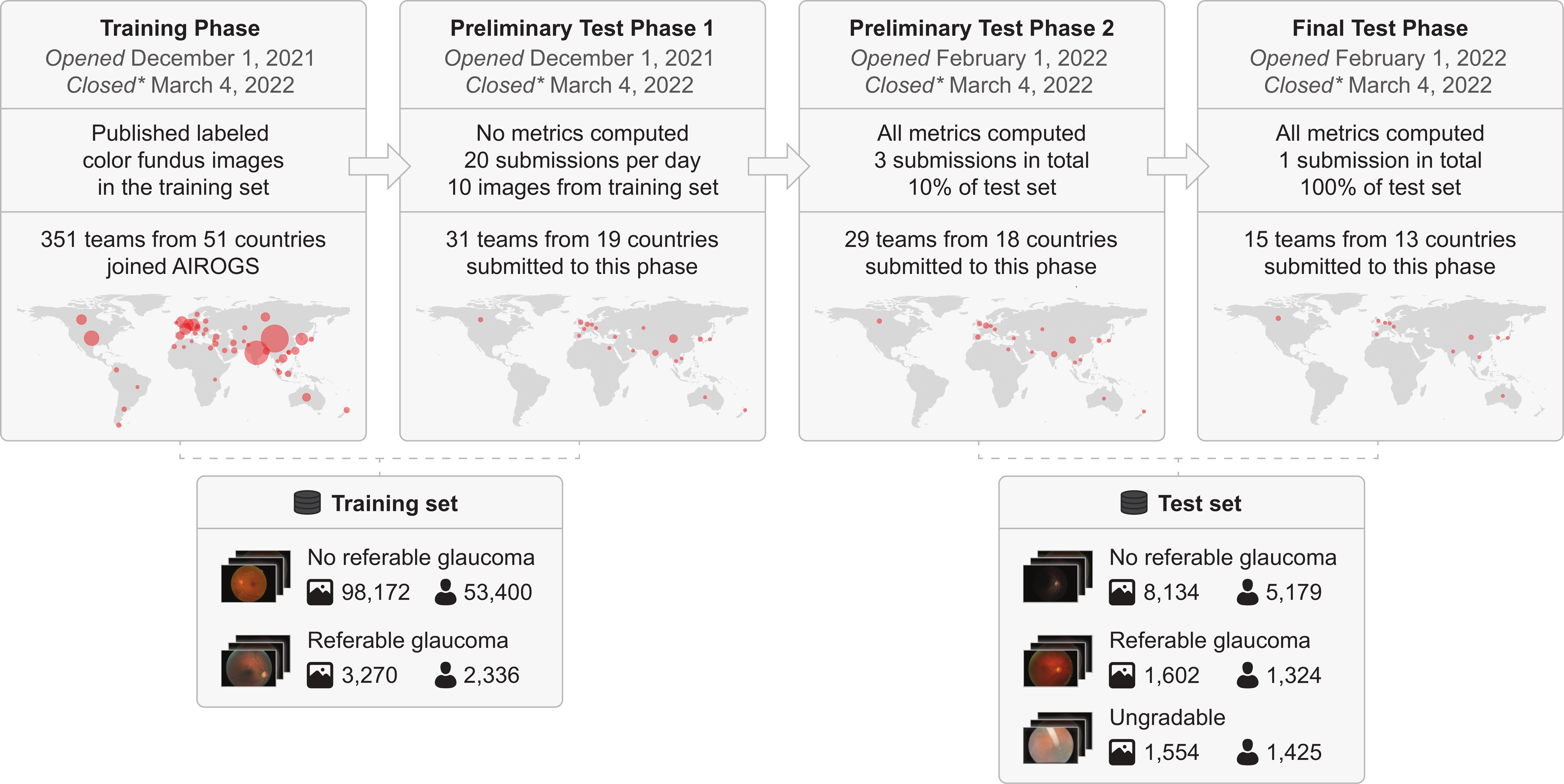}}
\caption{Overview of all phases in the AIROGS challenge. A world map is shown for each phase that indicates with red circles from which countries the teams that participated in that phase originated. A circle is shown for each country from which at least one team participated and its size represents the number of teams that joined from that country. The relevant subset of the AIROGS dataset for each phase is shown at the bottom of the figure. *All phases reopened for new submissions after the winning teams were announced.}
\label{fig:main_figure}
\end{figure*}


\subsection{The Rotterdam EyePACS AIROGS dataset}
\label{sec:datasets:challenge_dataset}
The Rotterdam EyePACS AIROGS dataset contains 112,732 CFPs from 60,071 subjects and 500 different sites with a heterogeneous ethnicity. The images were originally acquired for a diabetic retinopathy screening program \cite{Cuad09}. For grading of the CFPs, all graders were trained and then selected for this task using the European Optic Disc Assessment Trial (EODAT) \cite{Reus10}, containing 110 steroscopic optic nerve photographs, in which all glaucomatous eyes had reproducible visual field defects on standard automated perimetry. 90 experienced ophthalmologists and optometrists were examined and those who scored at least 85\% overall accuracy and 92\% specificity were selected to label images for the present study. Eventually, 30 out of 90 candidates passed. 


For each eye, three images were taken by the camera operators to reduce the number of ungradable eyes. When labeling the images, graders classified one eye at a time. The labeling tool first presented the first CFP for each eye, upon when graders could choose from the options "Referable glaucoma" (RG), "No referable glaucoma" (NRG) or "Ungradable" (U). If a grader selected U for the first image, the tool showed the consecutive CFP. The third image was presented if the second image was also deemed U. Each eye was scored by two separate graders, who were both unaware about the identity of the other grader. If the two graders agreed on the label of a CFP, this became the final label. If they disagreed, the image was scored by one of the glaucoma specialists who passed the EODAT test with at least 95\% accuracy. The final label was then based on his judgement.

The graders were instructed to select RG if they found glaucomatous signs which they expected to be associated with visual field defects on standard automated perimetry. The signs that could be selected were "appearance neuroretinal rim superiorly", "appearance neuroretinal rim inferiorly", "baring of the circumlinear vessel superiorly", "baring of the circumlinear vessel inferiorly", "disc hemorrhage(s)", "retinal nerve fibre layer defect superiorly", "retinal nerve fibre layer defect inferiorly", "nasalisation (nasal displacement) of the vessel trunk", "laminar dots" and "large cup". If the graders did not expect any glaucomatous visual field defects, NRG was to be selected, ignoring comorbidities such as age-related macular degeneration and diabetic retinopathy. If there was not enough information visible in the CFP to decide between RG and NRG, graders were instructed to select U. 

The graders were not only evaluated at the start, but they were periodically monitored during the grading process, as well. If their sensitivity or specificity dropped below 80\% or 95\%, respectively, they were removed from the study and all images they labeled were re-graded by any of the remaining graders. In case a grader wrongly classified a CFP as U, while its final label was NRG or RG, their specificity or sensitivity went down, respectively. In the end, 20 graders remained.

Out of the three CFPs that were available for each eye, we only included the RG or NRG photograph in the dataset if it was available. Otherwise, only one of the U photographs was used. We split the data into a training set of 101,442 CFPs and a test set of 11,290 CFPs, ensuring that data from patients in the training set was not in the test set. We randomly sampled patients when making the split, oversampling patients with ungradable and RG CFPs for the test set, such that approximately 1,600 RG and 1,600 U photographs ended up in the test set. Since we were interested in AI solutions that can identify ungradable data without training on ungradable data, we left out all U photographs that ended up in the AIROGS training set. Table \ref{table:dataset_stats} shows statistics about RG, NRG and U prevalence, age, sites and cameras for the full dataset, the training set and the test set. Approval from the Institutional Review Board of the Rotterdam Eye Hospital was obtained to conduct this research.

\subsection{External datasets}

The participants uploaded their trained algorithms, rather than a file with predictions on our test set, to our challenge platform. This enabled us to reuse the developed models on external data after the challenge ended. To evaluate model generalization and to demonstrate this reusability, we applied all trained algorithms to three external datasets: Retinal Fundus Glaucoma Challenge (REFUGE) \cite{Orla20}, Glaucoma grAding from Multi-Modality imAges (GAMMA) \cite{Wu22} and Diabetic Retinopathy Image Database (DRIMDB) \cite{Sevi14}. The former two are datasets with positive and negative glaucoma CFPs, which we used for externally evaluating the screening performance. We used the latter dataset, which contained different types of ungradable images, to externally evaluate the robustness.

The REFUGE test set contained 400 CFPs, of which 40 CFPs showed glaucoma and 360 CFPs did not. The definition of glaucoma was glaucomatous damage in the optic nerve head area and reproducible glaucomatous visual field defects, which is similar to our definition of glaucoma described earlier \cite{Orla20}.

The GAMMA dataset is a multi-modal dataset with optical coherence tomography scans and CFPs for each eye. We used the CFP data from the 100-sample training set as only that subset of the GAMMA dataset had publicly available labels. We defined positive glaucoma as the union of the intermediate and advanced glaucoma stages. These stages were defined using the mean deviation (MD) from the visual field reports as follows: an MD between than -6 dB and -12 dB for the intermediate stage and an MD worse than -12 dB for the advanced stage \cite{Wu22}. This resulted in 50 negative and 50 positive glaucoma samples.

DRIMDB is a dataset with 125 "Good" CFPs, 69 "Bad" CFPs and 22 "Outlier" CFPs. As the "Good" class was not necessarily defined as gradable for glaucoma, we manually confirmed that in all "Good" CFPs the optic disc (OD) was well visible. We defined both "Bad" and "Outlier" classes to be ungradable and the "Good" class to be gradable. "Bad" images were CFPs with low image quality and "Outlier" images were non-retinal images.

\section{Challenge setup}
The AIROGS challenge consisted of four phases (see Fig. \ref{fig:main_figure}). The \textit{Training Phase} opened on the 1\textsuperscript{st} of December 2021 and closed on the 4\textsuperscript{th} of March 2022, providing the participants with approximately three months to develop their solutions. At the start of this phase, the training set was released and has since been available for download under the \mbox{CC BY-NC-ND licence} on Zenodo\footnote{https://zenodo.org/record/5793241}.

To ensure fair competition and to encourage the development of inherent robustness mechanisms, teams were not permitted to use additional fundus image training data, including weights pre-trained on fundus image data or in pre-processing steps such as OD segmentation. Manually labeling the challenge data and using the resulting annotations during training was allowed.

To test the algorithms developed by participants, they needed to wrap their trained algorithm in a Docker\footnote{https://docker.com} container and submit it to our challenge platform. This allows the submitted algorithms to be run on data that is not directly accessible by the participating teams. Example code for generating such a containerized submission can be found on GitHub\footnote{https://github.com/qurAI-amsterdam/airogs-example-algorithm}. \textit{Preliminary Test Phase 1} opened and closed simultaneously with the \textit{Training phase} and served as a check for whether the submitted algorithms could be run on the challenge platform and produced the output in the expected format. Algorithms were tested on 10 images from the training set for this check. All algorithms were executed on the challenge platform using an NVIDIA T4 GPU (16 GB VRAM) with 8 CPUs (32 GB RAM). 

The test set was and still is closed, meaning the image data and the labels are private and cannot be downloaded. \textit{Preliminary Test Phase 2} opened on the 1\textsuperscript{st} of February 2022 and we allowed three submissions per team to this phase, as it used 10\% of the test set for evaluation. All challenge metrics were also computed and reported back to the participants. The \textit{Final Test Phase} opened simultaneously with \textit{Preliminary Test Phase 2}, but algorithms were tested on 100\% of the test data and only one submission per team was allowed. The challenge metrics computed for this phase were used for the final team ranking.

The algorithms were expected to produce four outputs, of which two were related to glaucoma screening performance (i.e., image classification of RG and NRG) and the other two to robustness (i.e., the identification of U). The glaucoma screening outputs were a likelihood score for RG ($O_1$) and a binary decision for RG ($O_2$, positive if RG and negative if NRG). The ungradability outputs were a binary decision on whether the image is ungradable ($O_3$, positive if ungradable and negative if ungradable) and a non-thresholded scalar value that is positively correlated with the likelihood for ungradability (e.g. the entropy of a probability vector produced by a machine learning model or the variance of an ensemble) ($O_4$). Output $O_2$ was not used in the evaluation pipeline for the challenge leaderboard, but it was requested by the challenge organizers for further analysis.

The evaluation was also based on the two aspects of screening performance and robustness, with two metrics per aspect. Screening performance was evaluated using the standardized partial area under the receiver operating characteristic curve \cite{Mccl89} (90-100\% specificity) for RG ($pAUC_S$), and the sensitivity at 95\% specificity ($SE@95SP_S$). These metrics were based on these specificity ranges, as a high specificity is required for cost-effective glaucoma screening due to its relatively low prevalence \cite{Vaah07,Vrie12}. $pAUC_S$ and $SE@95SP_S$ are both based on output $O_1$. For evaluating the robustness, we determined the model's agreement with the human reference on ungradability using Cohen's kappa score ($\kappa_U$), calculated using output $O_3$. Furthermore, we calculated the area under the receiver operator characteristic curve using the human reference for ungradability as the true labels and output $O_4$ as the target scores ($AUC_U$).

To determine the final ranking, we first ranked all participants on the four individual metrics $pAUC_S$, $SE@95SP_S$, $\kappa_U$, and $AUC_U$ resulting in the rankings $R_{pAUC_S}$, $R_{SE@95SP_S}$, $R_{\kappa_U}$, and $R_{AUC_U}$, respectively. The final score $S_{final}$ was then calculated as the mean of those ranking:

\begin{equation}
    S_{final} = \frac{R_{pAUC_S} + R_{SE@95SP_S} + R{\kappa_U} + R_{AUC_U}}{4}.
\end{equation}

The final ranking (later also referred to as \textit{Mean position}), was based on $S_{final}$, where a lower value for $S_{final}$ resulted in a higher ranking.

We calculated 95\% confidence intervals (CIs) with non-parametric bootstrapping using 1000 iterations \cite{Rutt20}. The code for evaluating submissions can be found on GitHub\footnote{https://github.com/qurAI-amsterdam/airogs-evaluation}. The performance of human graders was calculated by comparing the labels given by the individual graders (excluding the two glaucoma specialists, since the final labels were equal to their decision in case of disagreement) to the final labels as defined in Section \ref{sec:datasets:challenge_dataset}. To compute the performance of all human graders combined, each image was weighted equally in the calculation of the metrics. We also evaluated ensembles of participating algorithms, which were generated by averaging the outputs of these algorithms.

\section{Participating Methods}
\newcommand{\rot}[1]{\rotatebox[origin=l]{90}{#1}}

\newcolumntype{?}{!{\color{white}\vrule width .75pt}}
\newcolumntype{S}{@{\hskip 0.03in}}
\definecolor{lightgray}{rgb}{.9,.9,.9}

\begin{table*}[ht]
    \centering
    \setlength\tabcolsep{0.1pt}
    \caption{Method overview from all participating teams for the screening task. OD = optic disc.}
    \resizebox{\textwidth}{!}{\begin{tabular}{c?l?Sc?c?c?c?c?c?c?c?c?c?c?c?c?c?c?Sc?S>{\centering\arraybackslash}p{.35cm}?>{\centering\arraybackslash}p{.35cm}?c?Sc?c?S>{\centering\arraybackslash}p{.28cm}?>{\centering\arraybackslash}p{.28cm}?cS?c?c?c?cS?c?c?cS?>{\centering\arraybackslash}p{.5cm}?cS?c?c?c?c?c?c?c?c?c?c?c?c?c?c?c}

        \toprule
        \multirow{3}{*}[-7.5em]{\#} & \multirow{3}{*}[-7.5em]{Team} & \multicolumn{42}{c}{Screening} \\
        \cmidrule(lr){3-50}
        
        & & \multicolumn{15}{c}{Architecture} & & \multicolumn{3}{c}{\makecell{Pre-\\processing}} & & & \multicolumn{3}{c}{\makecell{Loss\\function}} & \multicolumn{4}{c}{Optimizer} & \multicolumn{3}{c}{\makecell{Pre-\\training}} & \multicolumn{2}{c}{\makecell{Class\\imbalance\\solution}} & \multicolumn{15}{c}{\makecell{Data augmentation\\during training}} \\
        
        \cmidrule(lr){3-17} \cmidrule(lr){19-21} \cmidrule(lr){24-26} \cmidrule(lr){27-30} \cmidrule(lr){31-33}  \cmidrule(lr){34-35} \cmidrule(lr){36-50}
        
        & & \rot{Vision transformer} & \rot{Swin-Transformer-B} & \rot{EfficientNet} & \rot{DeiT-S} & \rot{MobileNet} & \rot{ResNet} & \rot{ResNet-RS} & \rot{SeResNeXt} & \rot{VGG} & \rot{DenseNet} & \rot{Inception-V3} & \rot{ConvNeXt} & \rot{SeNet} & \rot{SeResNet} & \rot{Inception-ResNet-v2} & \rot{Ensemble} & \rot{Crop field-of-view} & \rot{Crop OD} & \rot{Histogram equalization} & \rot{\makecell[l]{Number of manually \\ labeled ODs}} & \rot{\makecell[l]{Input size of \\ classification model}} & \rot{Cross-entropy} & \rot{Focal loss} & \rot{Kullback-Leibler} & \rot{Adam \cite{King14}} & \rot{AdamW \cite{Losh17}} & \rot{SGD} & \rot{SAM \cite{Fore20}} & \rot{ImageNet} & \rot{Masked auto-encoder \cite{He22}} & \rot{Noisy student \cite{Xie20}} & \rot{Balanced sampling} & \rot{Weighted loss} & \rot{Brightness modifications} & \rot{Color jitter} & \rot{Elastic deformations} & \rot{Flipping} & \rot{Bluring} & \rot{Rotation} & \rot{Translation} & \rot{Zooming} & \rot{Shearing} & \rot{Quality reduction} & \rot{Sharpening} & \rot{Erasing} & \rot{AutoAugment \cite{Cubu18}} & \rot{RandAugment \cite{Cubu20}} & \rot{TrivialAugment \cite{Mull21}} \\

        \midrule
        
        \rowcolor{lightgray}
1 & PUMCH-eye & \checkmark &  &  &  &  &  &  &  &  &  &  &  &  &  &  &  &  & \checkmark &  & 40 & 384\textsuperscript{2} & \checkmark &  &  &  & \checkmark &  &  &  & \checkmark &  &  &  &  &  &  &  &  &  &  &  &  &  &  &  & \checkmark &  & 
 \\ 
2 & RWTH-CuP &  & \checkmark & \checkmark & \checkmark &  &  &  &  &  &  &  &  &  &  &  & \checkmark &  & \checkmark & \checkmark & 3,221 & 224\textsuperscript{2} & \checkmark &  &  & \checkmark & \checkmark &  &  & \checkmark &  &  & \checkmark &  &  &  & \checkmark & \checkmark &  &  &  &  &  & \checkmark &  &  &  &  & 
 \\ 
\rowcolor{lightgray}
2 & Eyelab & \checkmark &  &  &  &  &  &  &  &  &  &  &  &  &  &  &  &  & \checkmark &  & 1,500 & 384\textsuperscript{2} & \checkmark &  &  &  &  & \checkmark &  &  &  &  & \checkmark &  & \checkmark &  &  & \checkmark &  & \checkmark & \checkmark & \checkmark & \checkmark &  &  & \checkmark &  &  & 
 \\ 
4 & Tien &  &  & \checkmark &  &  &  &  &  &  & \checkmark &  &  &  &  &  & \checkmark & \checkmark &  &  &  & 1,024\textsuperscript{2} & \checkmark &  &  & \checkmark &  &  &  &  &  &  & \checkmark &  &  &  &  & \checkmark &  & \checkmark &  & \checkmark &  &  &  &  &  &  & 
 \\ 
\rowcolor{lightgray}
5 & UPF+AIML &  &  &  &  & \checkmark &  &  &  &  &  &  &  &  &  &  & \checkmark &  &  &  &  & 512\textsuperscript{2} & \checkmark &  &  &  & \checkmark &  & \checkmark & \checkmark &  &  &  &  &  &  &  &  &  &  &  &  &  &  &  &  &  &  & \checkmark
 \\ 
6 & FMS-CETCV &  &  &  &  &  & \checkmark &  &  &  &  &  &  &  &  &  &  &  &  &  &  & 512\textsuperscript{2} &  & \checkmark &  &  &  & \checkmark &  & \checkmark &  &  &  &  &  &  &  & \checkmark &  &  &  &  &  &  &  &  &  &  & 
 \\ 
\rowcolor{lightgray}
7 & ICT\_HCI &  &  &  &  &  & \checkmark &  &  &  &  &  &  &  &  &  &  & \checkmark &  &  &  & 512\textsuperscript{2} & \checkmark &  &  & \checkmark &  &  &  & \checkmark &  &  & \checkmark &  &  &  &  &  &  &  &  &  &  &  &  &  &  & \checkmark & 
 \\ 
8 & SK &  &  &  &  &  &  & \checkmark &  &  &  &  &  &  &  &  & \checkmark & \checkmark &  &  &  & 256\textsuperscript{2} & \checkmark &  &  & \checkmark &  &  &  & \checkmark &  &  &  & \checkmark &  & \checkmark &  &  & \checkmark & \checkmark & \checkmark & \checkmark &  &  &  &  &  &  & 
 \\ 
\rowcolor{lightgray}
9 & SACM &  &  &  &  &  &  &  & \checkmark & \checkmark & \checkmark & \checkmark &  &  &  &  & \checkmark &  & \checkmark &  & 735 & 120\textsuperscript{2}/224\textsuperscript{2} & \checkmark &  &  &  & \checkmark & \checkmark &  & \checkmark &  &  &  & \checkmark &  &  &  &  &  &  &  &  &  &  &  &  &  &  & 
 \\ 
10 & UPRetina-UR &  &  & \checkmark &  &  &  & \checkmark &  &  &  &  &  &  &  &  &  & \checkmark &  &  &  & 512\textsuperscript{2} &  & \checkmark &  & \checkmark &  &  &  & \checkmark &  &  & \checkmark &  &  &  &  &  &  &  &  &  &  &  &  &  &  & \checkmark & 
 \\ 
\rowcolor{lightgray}
11 & OPTIMATeam &  &  &  &  &  &  &  &  &  &  & \checkmark &  &  &  &  &  &  &  &  &  & 224\textsuperscript{2} &  &  & \checkmark & \checkmark &  &  &  & \checkmark &  &  & \checkmark &  & \checkmark &  &  & \checkmark & \checkmark & \checkmark & \checkmark & \checkmark &  &  &  &  &  &  & 
 \\ 
11 & MA &  &  & \checkmark &  &  &  &  & \checkmark &  & \checkmark &  &  & \checkmark & \checkmark & \checkmark &  & \checkmark &  &  &  & 512\textsuperscript{2} & \checkmark &  &  & \checkmark &  &  &  & \checkmark &  &  & \checkmark &  & \checkmark &  &  & \checkmark & \checkmark & \checkmark & \checkmark & \checkmark &  &  & \checkmark &  &  &  & 
 \\ 
\rowcolor{lightgray}
13 & YC &  &  &  &  &  &  &  &  &  & \checkmark &  &  &  &  &  & \checkmark &  & \checkmark & \checkmark & 101 & 608\textsuperscript{2} & \checkmark &  &  &  &  & \checkmark &  &  &  & \checkmark &  & \checkmark & \checkmark &  &  & \checkmark & \checkmark & \checkmark & \checkmark & \checkmark &  &  &  &  &  &  & 
 \\ 
14 & Mirazzak & \checkmark &  &  &  &  &  &  &  &  &  &  & \checkmark &  &  &  & \checkmark &  &  &  &  & 224\textsuperscript{2}/384\textsuperscript{2} & \checkmark &  &  & \checkmark &  &  &  & \checkmark &  &  &  & \checkmark &  &  &  & \checkmark &  & \checkmark &  &  &  &  &  &  &  &  & 
 \\

        \bottomrule
    \end{tabular}}
\label{table:screening_table}
\end{table*}

\begin{table*}[t]
    \centering
    \caption{Method overview from all participating teams for the ungradability task and the deep learning frameworks they used. OD = optic disc. AE = autoencoder. VAE = variational autoencoder. rec. error = reconstruction error. OOD = out-of-distribution.}
    \setlength\tabcolsep{2.0pt}
    \resizebox{\textwidth}{!}{\begin{tabular}{c?l@{\hskip 0.06in}?l?c?c?c@{\hskip 0.06in}?c}
        \toprule
        \multirow{2}{*}[-2.2em]{\#} & \multirow{2}{*}[-2.2em]{Team} & \multicolumn{4}{c}{Robustness} & General \\
        \cmidrule(lr){3-6} \cmidrule(lr){7-7}
        
        & & \multicolumn{1}{c}{Method} & \makecell{OD detection \\ for uncertainty \\ or confidence \\ estimation} & \makecell{Threshold based on \\ manual identification of \\ low quality images in \\ the development set}  & \makecell{Input size for \\ ungradability \\ approach} & \makecell{Deep \\ learning \\ framework} \\

        \midrule
        
        \rowcolor{lightgray}
1 & PUMCH-eye & Vessel and OD segmentation & \checkmark & \checkmark & 384\textsuperscript{2} & PyTorch
 \\ 
2 & RWTH-CuP & OD detection confidence & \checkmark & \checkmark & 224\textsuperscript{2} & PyTorch
 \\ 
\rowcolor{lightgray}
2 & Eyelab & OD presence detection & \checkmark & \checkmark & 384\textsuperscript{2} & PyTorch
 \\ 
4 & Tien & AE rec. error + probability classification model &  &  & 1,024\textsuperscript{2} & PyTorch
 \\ 
\rowcolor{lightgray}
5 & UPF+AIML & Synthetic image degradations &  & \checkmark & 512\textsuperscript{2} & PyTorch
 \\ 
6 & FMS-CETCV & Interpolated gaussian descriptor &  &  & 256\textsuperscript{2} & PyTorch
 \\ 
\rowcolor{lightgray}
7 & ICT\_HCI & Probability classification model &  & \checkmark & 512\textsuperscript{2} & PyTorch
 \\ 
8 & SK & Energy-based OOD + activation rectification &  &  & 256\textsuperscript{2} & PyTorch+Lightning
 \\ 
\rowcolor{lightgray}
9 & SACM & AE and VAE rec. error + OD detection confidence & \checkmark & \checkmark & 288\textsuperscript{2} & PyTorch
 \\ 
10 & UPRetina-UR & Test-time augmentation + probability classification model &  &  & 512\textsuperscript{2} & PyTorch+fast.ai
 \\ 
\rowcolor{lightgray}
11 & OPTIMATeam & Deep Dirichlet uncertainty &  &  & 224\textsuperscript{2} & TensorFlow+Keras
 \\ 
11 & MA & Ensemble &  &  & 512\textsuperscript{2} & TensorFlow+Keras
 \\ 
\rowcolor{lightgray}
13 & YC & Monte-Carlo Dropout &  &  & 608\textsuperscript{2} & TensorFlow
 \\ 
14 & Mirazzak & Regret function &  &  & 224\textsuperscript{2}/384\textsuperscript{2} & PyTorch
 \\

        \bottomrule
    \end{tabular}}
\label{table:robustness_general_table}
\end{table*}
    
Fifteen teams submitted a working solution to the \textit{Final Test Phase}, of which one team did not opt-in to contribute to the current paper. In this section, we present the methods of the fourteen participating teams.
More extensive descripions are available on the AIROGS challenge website\footnote{https://airogs.grand-challenge.org/evaluation/final-test-phase/leaderboard/} and a selection of the participating methods were included in the ISBI challenge proceedings \cite{Wang22,Khad22,Arau22}. Table \ref{table:screening_table} and \ref{table:robustness_general_table} summarize the participating methods in a structured manner.

\subsection{PUMCH-eye \cite{Wang22}} 
The \textit{PUMCH-eye} team proposed an approach with five trained models in their workflow. The first model ($M_{disc}$) was a segmentation model with ResNet101-UperNet \cite{Ziao18} as the backbone that segmented the OD in the input CFP. For the development of this model, they manually labeled the OD in 40 images. In case $M_{disc}$ successfully detected the OD, they computed the center $c$ and the diameter $d$ of the segmentation to crop the input image around $c$ with size $3d$. This cropped image was then fed into a vision transformer \cite{Shar21} for the binary classification of RG and NRG. If the OD detection was unsuccessful, they fed the original input image to a different vision transformer for binary classification of RG and NRG.

The team also developed a vessel segmentation model with 40 images in which they manually annotated vessels ($M_{vessel}$). They trained a ResNet-18 ($R_{vessel}$) which took the output of $M_{vessel}$ as input data, using the first 500 images in the training dataset and 100 manually selected images in the training set with relatively poor image quality. This classfication model served as one of the inputs for ungradability classification. The second input was taken from $M_{disc}$. The ungradability likelihood output ($O_4$) was then defined as the output likelihood of the binary classification model $R_{vessel}$ (i.e., $O_{vessel}$ or, if the $M_{disc}$ could not detect an OD, as $R_{vessel} + 0.75$. $O_3$ was positive if $O_4$ was at least $0.95$ and negative otherwise.

\subsection{RWTH-CuP \cite{Khad22}} 
The \textit{RWTH-CuP} team proposed an approach with two steps consisting of cropping around the OD by employing a detection network, followed by an ensemble of transformers (Swin Transformer-B \cite{Liu21} and DeiT-S \cite{Touv21}) and convolutional neural networks (EfficientNet-B4 \cite{Tan19} and EfficientNetV2-M \cite{Tan21}) that classifies the cropped image. They manually labeled the OD and its environment in 3,221 CFPs to develop this detection network, for which they trained a YOLOv5 \cite{Joch20} object detector network.

For ungradability classification, the team used a hybrid approach. As the probability that an image is ungradable is high if the OD could be found by the object detector network, they employed the confidence score of the YOLOv5 detection model as one of the ungradability measures. To capture other ungradability causes, such as blurred depictions of the OD, they trained an additional classifier on a manually selected subset of the CFPs in the development set. The team considered the 4000 CFPs with the lowest confidence score of the object detector and manually selected 600 images that were assumed by the team to be very close to being classified as ungradable. They used another set of 2,000 high quality images to train an EfficientNet-B4 \cite{Tan19} ungradability classification model. $O_4$ was then defined as $(1-c) + g$, where $c$ is the object detection confidence and $g$ the output the ungradability classification model. The binary $O_3$ output value was determined using a cut-off manually determined by a medical doctor in 20,000 images from the development set for which $O_4$ was computed.

\subsection{Eyelab \cite{Aimy22}} 
The \textit{Eyelab} team employed a two-stage approach for glaucoma classification. The first step was to detect and crop the OD area and the second step was a vision transformer \cite{Doso20} that classified the cropped image from the first step. For the detection model, they trained a YOLOv5 \cite{Joch20} model using semi-automatically generated labels. Their method for ungradability detection was based on whether the optic disc detection model from the first step found an optic disc to be present.

\subsection{Tien \cite{Le22}} 
\textit{Tien} used an ensemble of an EfficientNet \cite{Tan19} and DenseNet \cite{Huan17} for the classification of RG and NRG. For the ungradability task, they used an autoencoder network and a blending engine. 
They used the the reconstruction error as a measure of the likelihood of ungradability. The higher the reconstruction error, the more likely it is that the image is ungradable. The blending engine fused the probability output from the binary classification model  as a weight factor to the reconstruction error. The highest weight was 1 (when the probability was 0.5) and the weight was lowest when the probability is certain (either 0 or 1).

\subsection{UPF+AIML \cite{Gald22}} 
Team \textit{UPF+AIML} trained two separate models, both based on the MobileNet-V2 \cite{Howa17} architecture for lightweight training, and both optimized with the Sharpness-Aware Minimization (SAM) \cite{Fore20} technique for better generalization. The first model was trained on the available training set for the screening task. The second model was tasked with identifying out-of-distribution data, i.e., ungradable images in this case. For this, ungradable images were simulated by applying four image transformations (brightness, gamma, saturation and blur) online to the data, with such a strength that they would destroy image content and turn images useless for diagnostic purposes. The ungradability detection model was trained on a mixture of gradable (sampled directly from the original training set) and ungradable (simulated). After training, this model was applied on the training set, where all images were expected to be gradable. The threshold that would classify 0.1\% of the training set as ungradable was selected for ungradabiltiy detection.

\subsection{FMS-CETCV \cite{Puth22}} 
The \textit{FMS-CETCV} team used a binary classifier with ResNet-50 as backbone for classifying RG and NRG. They used focal loss \cite{Lin17} to account for the class imbalance in the training set.

For the classification of ungradable images, they used a self-supervised learning approach, inspired by the work of Oza \textit{et al.} \cite{Oza18}, where a one-class classification method was presented for unsupervised anomaly detection. The one-class classifier builds a feature space by extracting the features of the training sample which contain only the positive samples (i.e., gradable images). They used an encoder with ResNet-18 as backbone, which is trained on the AIROGS test set that only contains gradable images. The feature space produced by this encoder is then used by a Gaussian anomaly classifier to distinguish gradable and ungradable images.

\subsection{ICT\_HCI \cite{Yang22}} 
Team \textit{ICT\_HCI} used ResNet-50 for their RG and NRG classifcation model. During inference, they made five random 512x512 crops of the image and then provided all crops separately to the model and get five scores. If the maximum of five scores was greater than 0.9, they let it be the output of the model, otherwise they took the mean of the five scores as the output of the model.

The team used the minimum class probability of the two classes RG and NRG as the likelihood for ungradability. If and only if the ungradability likelihood was greater than 0.1, they set $O_3$ to be positive.

\subsection{SK \cite{Kond22}} 
The \textit{SK} team employed ResNet-RS \cite{Bell21} for RG and NRG classification. They replaced the final linear layer of ResNet-RS with a single linear layer with two channel outputs.

For ungradability classification, they used an inference-time OOD energy-based method \cite{Liu20} combined with activation rectification \cite{Sun21}. The energy-based method uses a scoring function based on energy, instead of softmax, to discriminate in-distribution (ID) and OOD data. In activation rectification, the outsized activation of a few layers can be attenuated by rectifying the activations at an upper limit. After rectification, the output distributions for ID and OOD data become much more well-separated. It is based on the observation that the mean activation for ID data is well-behaved with a near-constant mean and standard deviation, and the mean activation for OOD data has significantly larger variations across units and is biased towards having sharp positive values.

\subsection{SACM \cite{Wang22a}} 
Team \textit{SACM} used the YOLOv5 \cite{Joch20} detection model to crop the optic disc in the CFP input image. In a semi-automated process, they manually labeled the locations of 735 optic discs and trained the detection model with 4,088 in total. The cropped image was then passed through an ensemble of classifiers (SeResNext-50 \cite{Hu18}, VGG-16 \cite{Simo14}, DenseNet-161 \cite{Huan17}, EfficientNet-B5 \cite{Tan19}, EfficientNet-B7 \cite{Tan19} and Inception-V3 \cite{Szeg16}) to make the final prediction. They also used test-time augmentation.

For the robustness task, they used the detection model confidence, an autoencoder, and a variational autoencoder (VAE) \cite{King13}. They combined these three aspects of their pipeline using this formula to achieve a final ungradability score for $O_4$ as $(1-c)\cdot s \cdot p_{autoencoder} \cdot p_{vae}$, where $c$ refers to the detection network confidence and $p_{autoencoder}$ and $p_{vae}$ refer to the mean squared error between the input and output of the autoencoder and VAE, respectively. 

\subsection{UPRetina-UR \cite{Hera22}} 
The \textit{UPRetina-UR} team used ResNet-RS-50 \cite{Bell21} for the classification of RG and NRG. They oversampled cases with RG during training to account for the class imbalance.

They employed a closed-set classification approach for the ungradability task based on the method proposed by Vaze \textit{et al.} \cite{Vaze21}. They applied test-time augmentation to obtain five predictions that are averaged to produce $O_4$.

\subsection{OPTIMATeam \cite{Arau22}} 
\textit{OPTIMATeam} used the first two blocks from the Inception-V3 \cite{Szeg16} network for the classification of RG and NRG. They only used these two blocks to reduce the receptive field size, which was necessary for their ungradability approach.

The ungradability approach was based on the direct \mbox{modeling} of the uncertainty following the evidential deep learning approach \cite{Demp68}. They used Deep Dirichlet uncertainty estimation as the ungradability score $O_4$. To set a threshold for getting a binary value for $O_3$ based on $O_4$, they made the assumption that diagnosis is only possible if the OD has enough image quality for diagnosis, as glaucoma main structural manifestation occurs in that region. They applied Grad-CAM \cite{Selv17} on the trained model for the screening task and occluded out the region where Grad-CAM was greater than 0.5. This allowed them to produce ID and OOD samples in their validation set, with which they computed the threshold for the binary ungradability decision. In particular, they constructed a receiver operating characteristic (ROC) curve using their values for $O_4$ with these ID and OOD samples. The ROC threshold where the sensitivity was 0.5 was set to calculate $O_3$.

\subsection{MA \cite{Arik22}} 
Team \textit{MA} used an ensemble of these twelve different architectures for the glaucoma screening task: SeNet-154 \cite{Hu18}, SeResNet-101 \cite{Hu18}, SeResNeXt-101 \cite{Hu18}, EfficientNet-B1 \cite{Tan19}, EfficientNet-B2 \cite{Tan19}, EfficientNet-B3 \cite{Tan19}, EfficientNet-B4 \cite{Tan19}, EfficientNet-B5 \cite{Tan19}, EfficientNet-B6 \cite{Tan19}, EfficientNet-B7 \cite{Tan19}, DenseNet-201 \cite{Huan17}, Inception-ResNet-v2 \cite{Szeg17}. The RG likelihood was computed by averaging the likelihoods of all respective models in the ensemble.

The ungradability output $O_4$ was the sum of the variances between all models in the ensemble for the positive and negative class probabilities. $O_3$ was positive if $O_4$ exceeded 0.2 and negative otherwise.

\subsection{YC \cite{Chan22}} 
The \textit{YC} team used two DenseNet-121 networks to classify RG and NRG in the CFPs. The first network was trained with the full CFP as input and the second network used a version of the CFP that was cropped around the optic disc as input. After the last convolutions of these networks, a fully connected layer with dropout was added. The outputs of these fully connected layers were then concatenated and used as the input to another fully connected layer with dropout, which was followed by the final layer of the network. For cropping the CFPs around the optic disc, they trained a U-Net \cite{Ronn15} with a DenseNet-121 \cite{Huan17} backbone. To train this segmentation network, they first roughly annotated the position of the optic disc in 101 CFPs in the training set. Subsequently, they generated reference segmentation maps using a probability density function of the multivariate normal distribution around the annotated optic disc position.

They used Monte-Carlo drop-out \cite{Gal16} with 20 predicted probabilties per image for the robustness task. Then they statistically tested a Wilcoxon one-sample test whether the mean of the predicted probabilities was 0.5. The team defined ungradability for predicting glaucoma as the logarithm of the p-value for the Wilcoxon test.

\subsection{Mirazzak \cite{Qayy22}} 
Team \textit{Mirazzak} used an ensemble of ConvNeXts \cite{Liu22} and a vision transformer for the screening performance task.

For ungradability task, they employed the \textit{regret} function, which was proposed by Bibas \textit{et al.} \cite{Biba21} as the generalization error of an explicit expression of the predictive normalized maximum likelihood learner.

If the value of \textit{regret} function was high, the samples were considered OOD and they marked them as ungradable.

\section{Results}

\begin{figure*}[!t]
\centerline{\includegraphics[width=.9\linewidth]{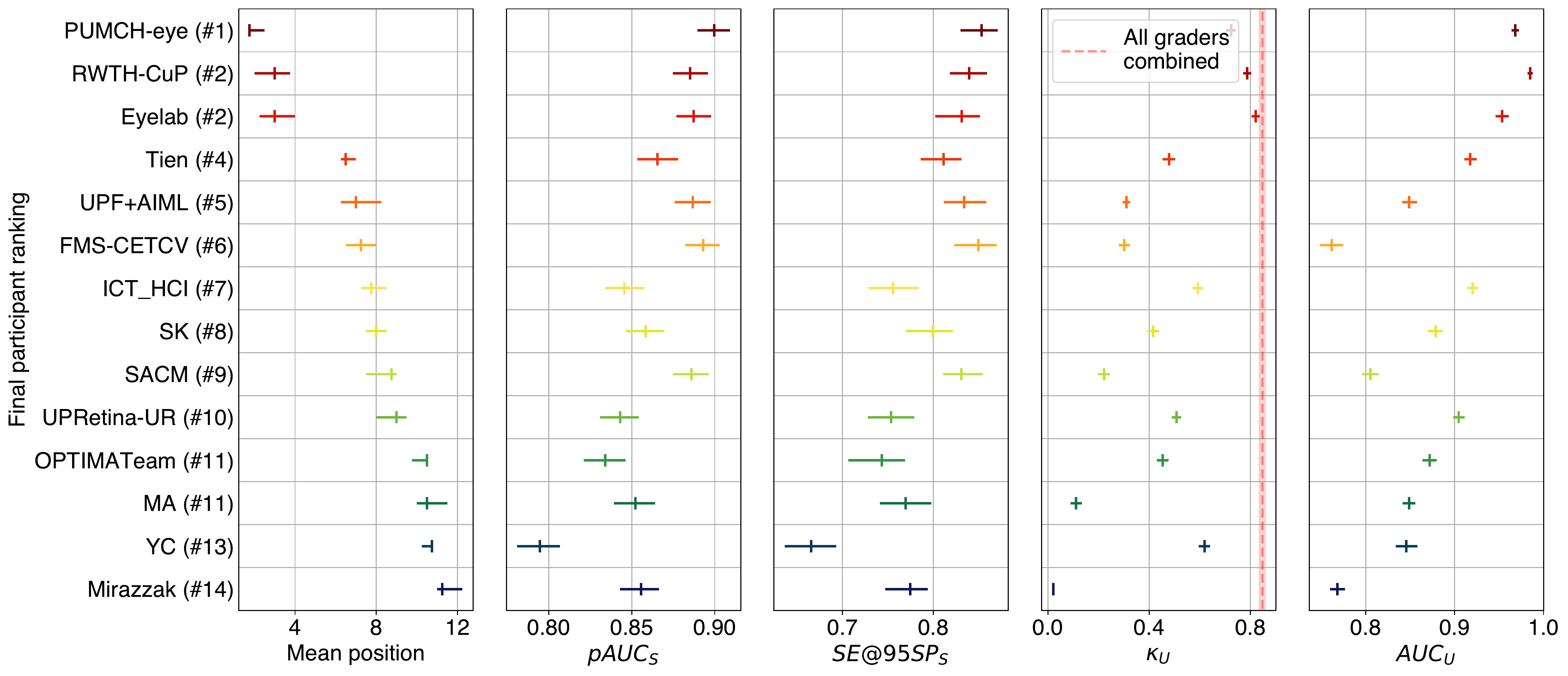}}
\caption{Final rankings of all participating teams. The teams are sorted by their final ranking and therefore also by their mean position. The mean position is shown in the left plot and the four challenge metrics are shown in the other four plots. The $\kappa_U$ of all human graders is indicated with a red dotted line. The width of the horizontal lines in all plots and the shaded area in the plot for $\kappa_U$ are 95\% CIs. We consistently use the same colors to refer to teams in other figures in this manuscript.}
\label{fig:all_metrics}
\end{figure*}

\begin{figure*}[!t]
\centering
\begin{subfigure}{0.49\textwidth}
    \includegraphics[width=\textwidth]{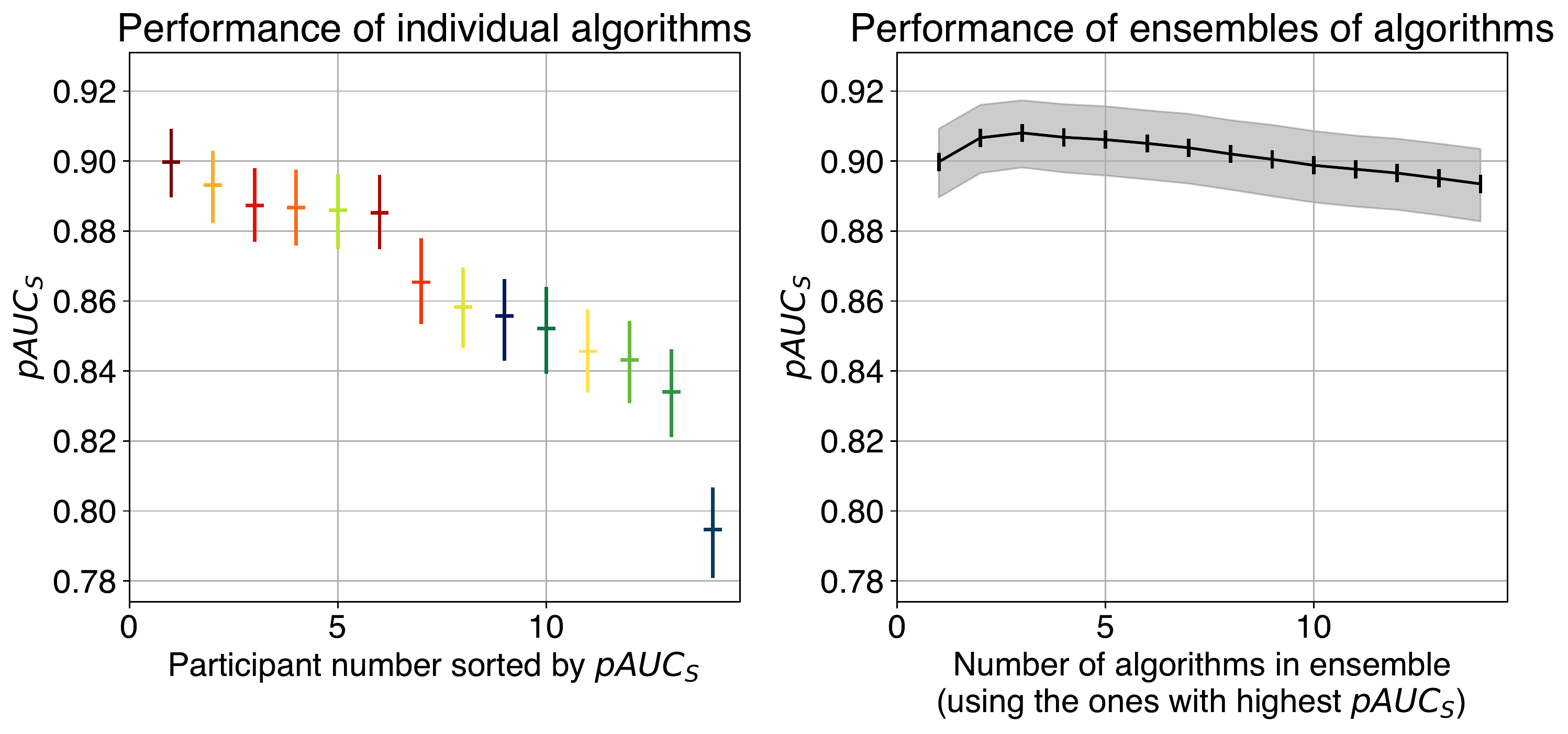}
    \caption{}
    \label{fig:ensembles_alpha}
\end{subfigure}
\hfill
\begin{subfigure}{0.49\textwidth}
    \includegraphics[width=\textwidth]{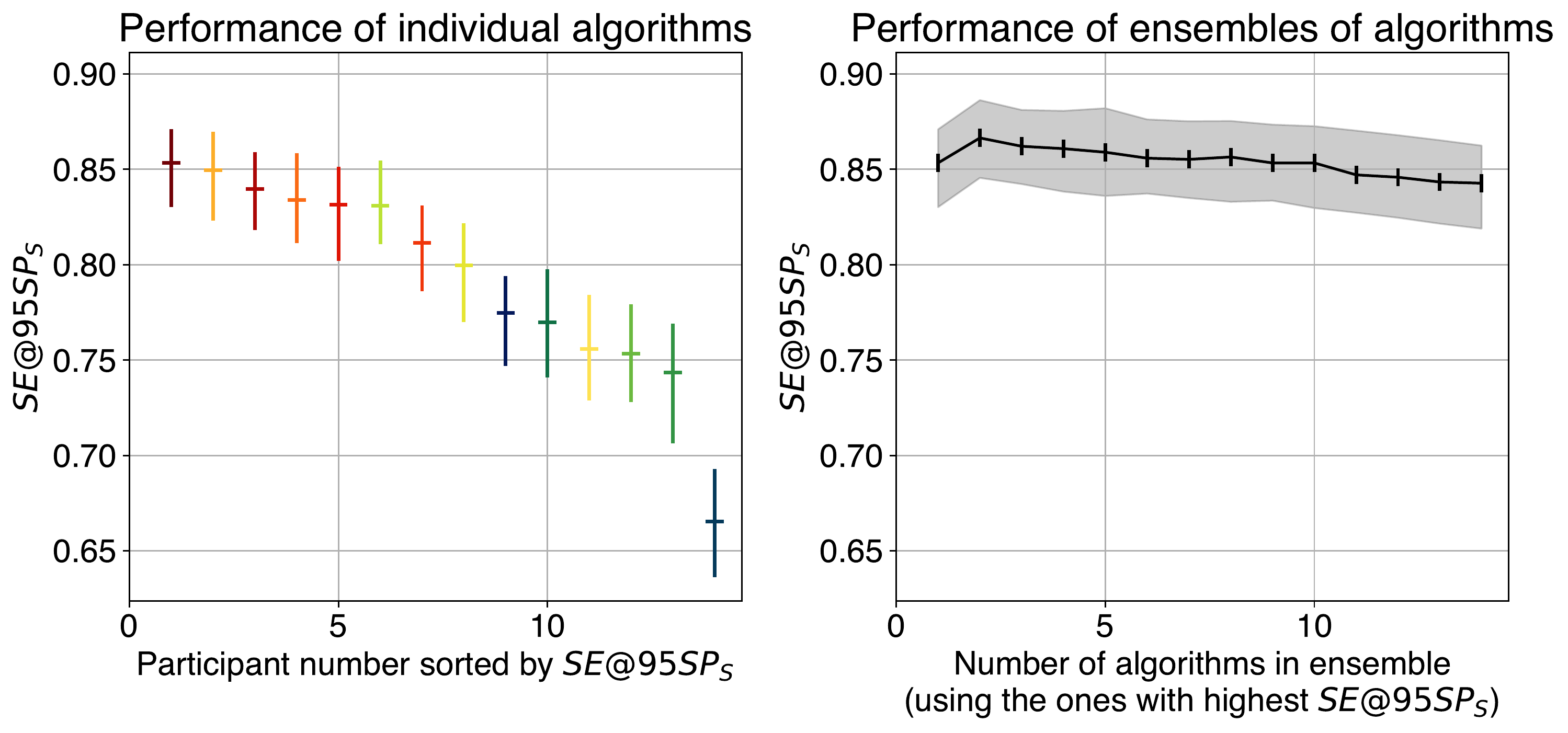}
    \caption{}
    \label{fig:ensembles_beta}
\end{subfigure}
\hfill
\begin{subfigure}{0.49\textwidth}
    \includegraphics[width=\textwidth]{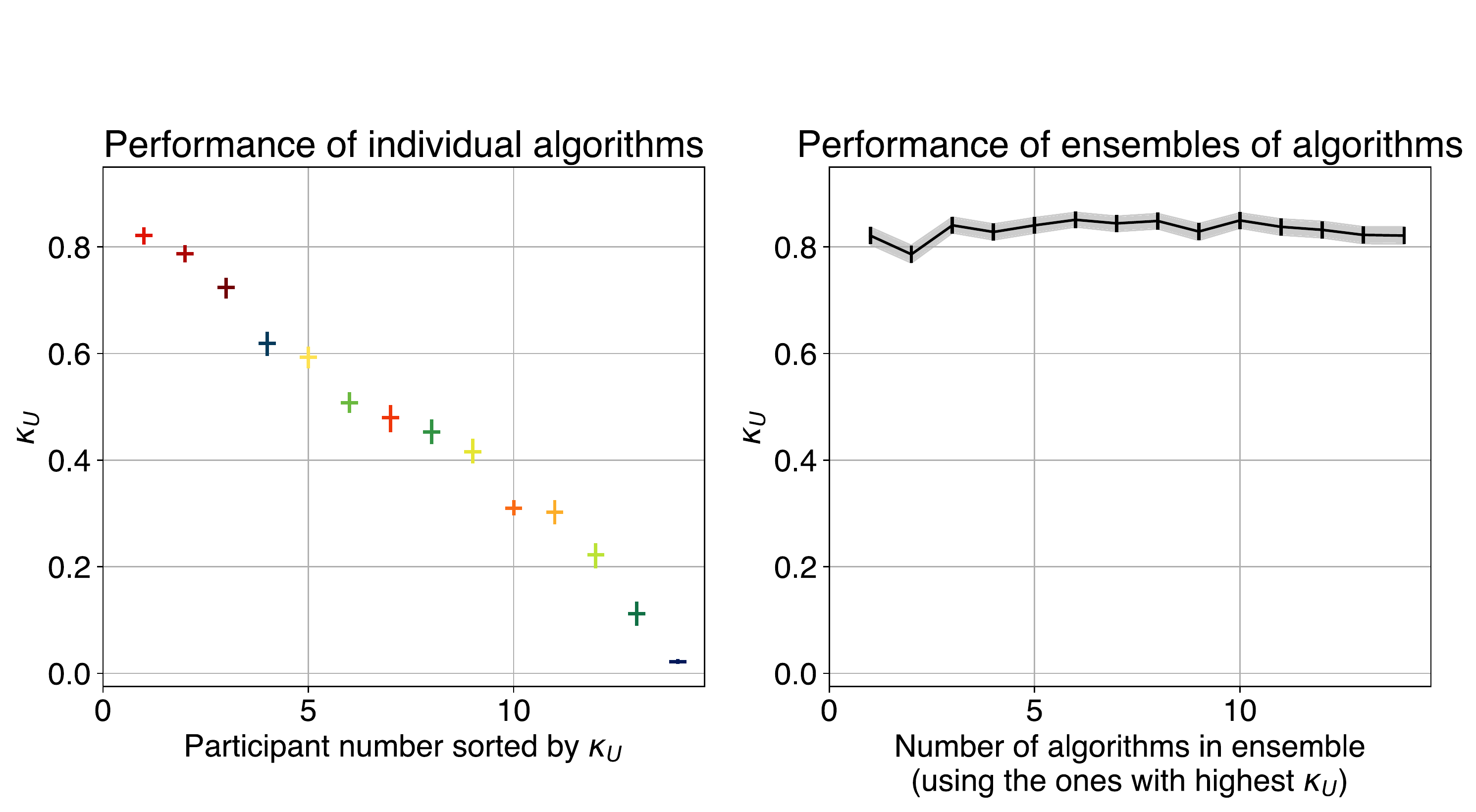}
    \caption{}
    \label{fig:ensembles_gamma}
\end{subfigure}
\hfill
\begin{subfigure}{0.49\textwidth}
    \includegraphics[width=\textwidth]{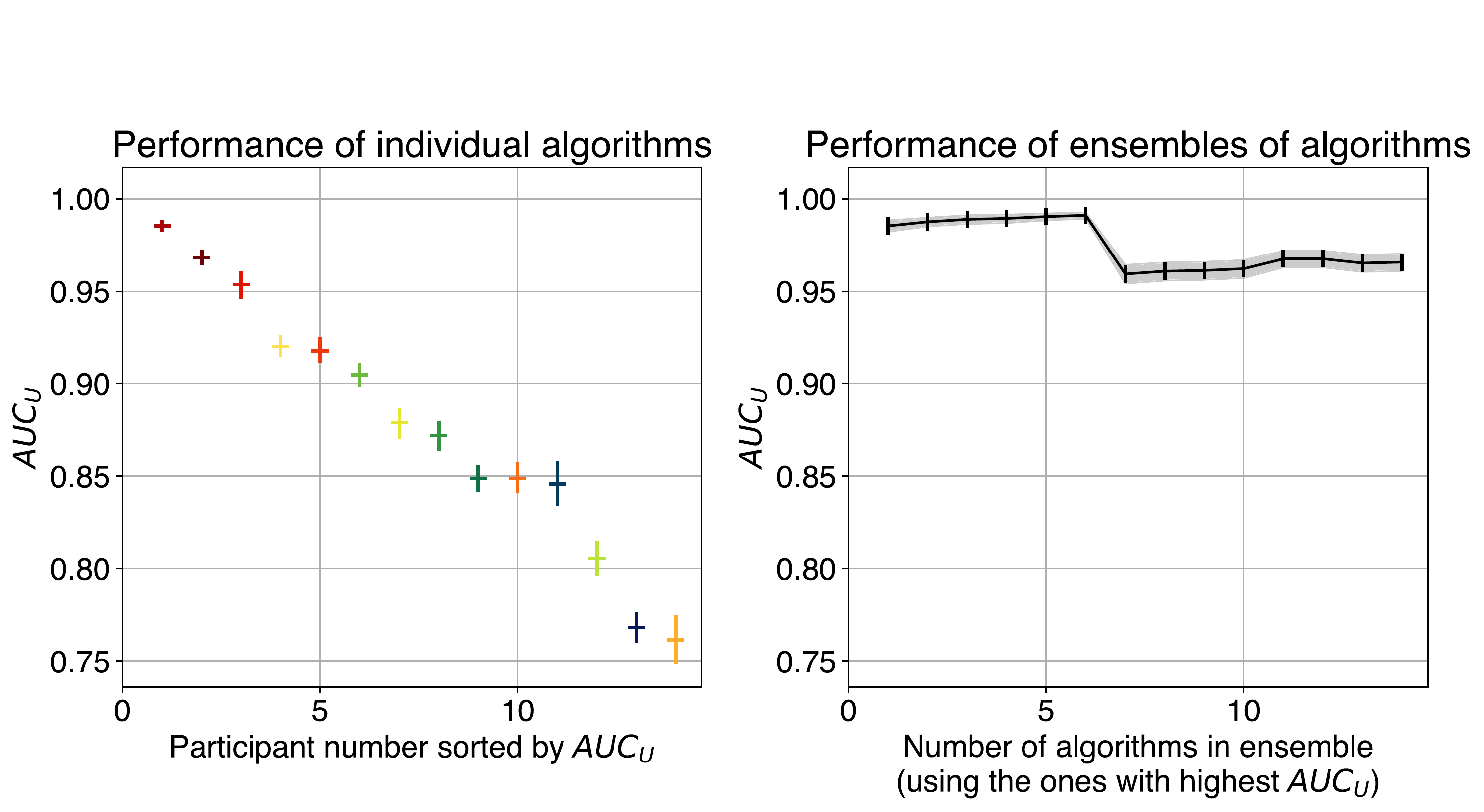}
    \caption{}
    \label{fig:ensembles_delta}
\end{subfigure}
        
\caption{The four challenge metrics (\protect\subref{fig:ensembles_alpha}) $pAUC_S$, (\protect\subref{fig:ensembles_beta}) $SE@95SP_S$, (\protect\subref{fig:ensembles_gamma}) $\kappa_U$, and (\protect\subref{fig:ensembles_delta}) $AUC_U$ for the ensembles generated by incrementally fusing one algorithm at a time. The algorithms were fused by averaging the outputs of all algorithms in the ensemble. The vertical lines in the left plot and the shaded areas in the right plots indicate 95\% CIs.}
\label{fig:ensembles}
\end{figure*}

\begin{figure*}[!t]
\centering
\begin{subfigure}{0.45\textwidth}
    \includegraphics[width=\textwidth]{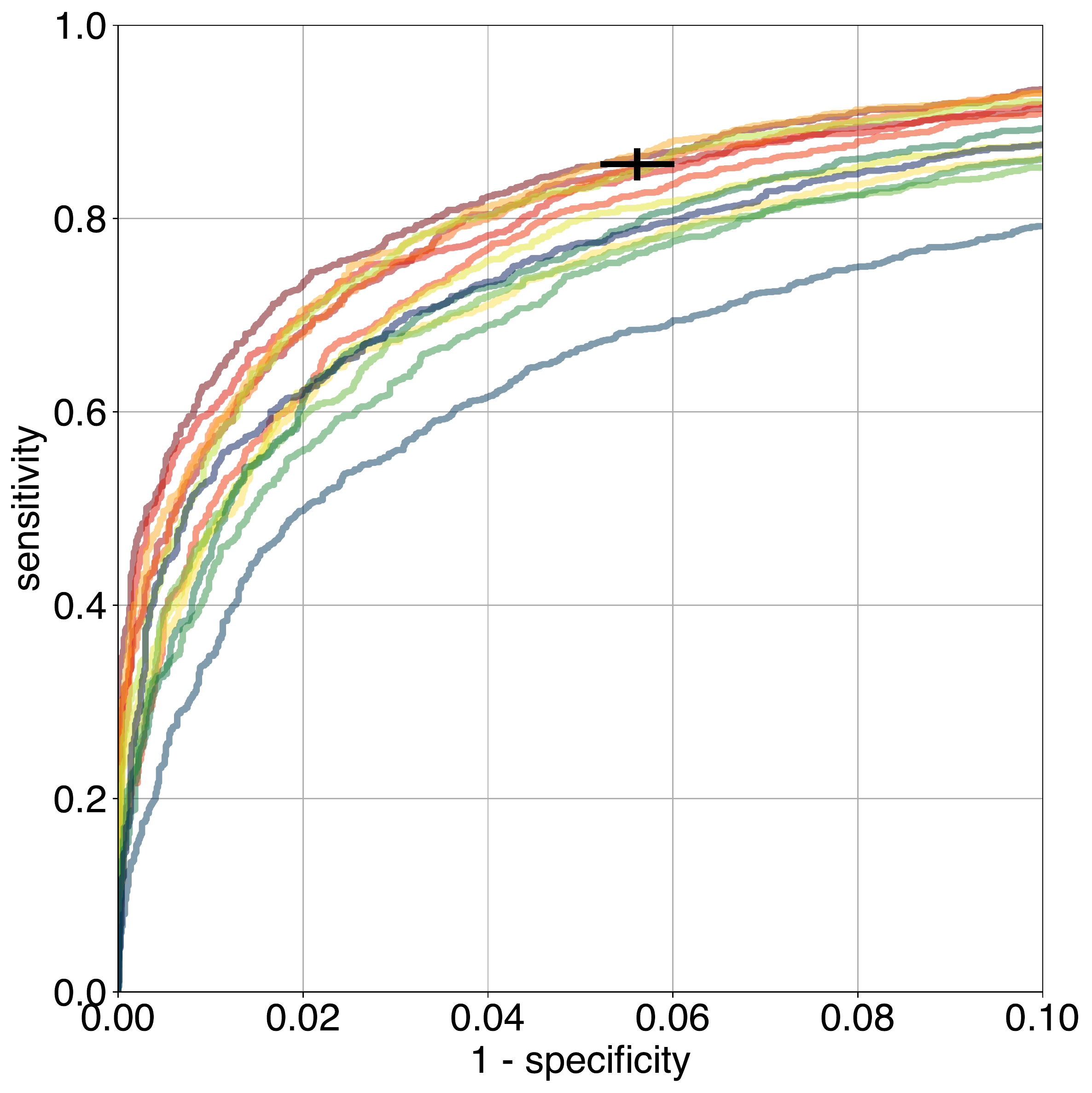}
    \caption{}
    \label{fig:roc_screening}
\end{subfigure}
\hspace{1cm}
\begin{subfigure}{0.45\textwidth}
    \includegraphics[width=\textwidth]{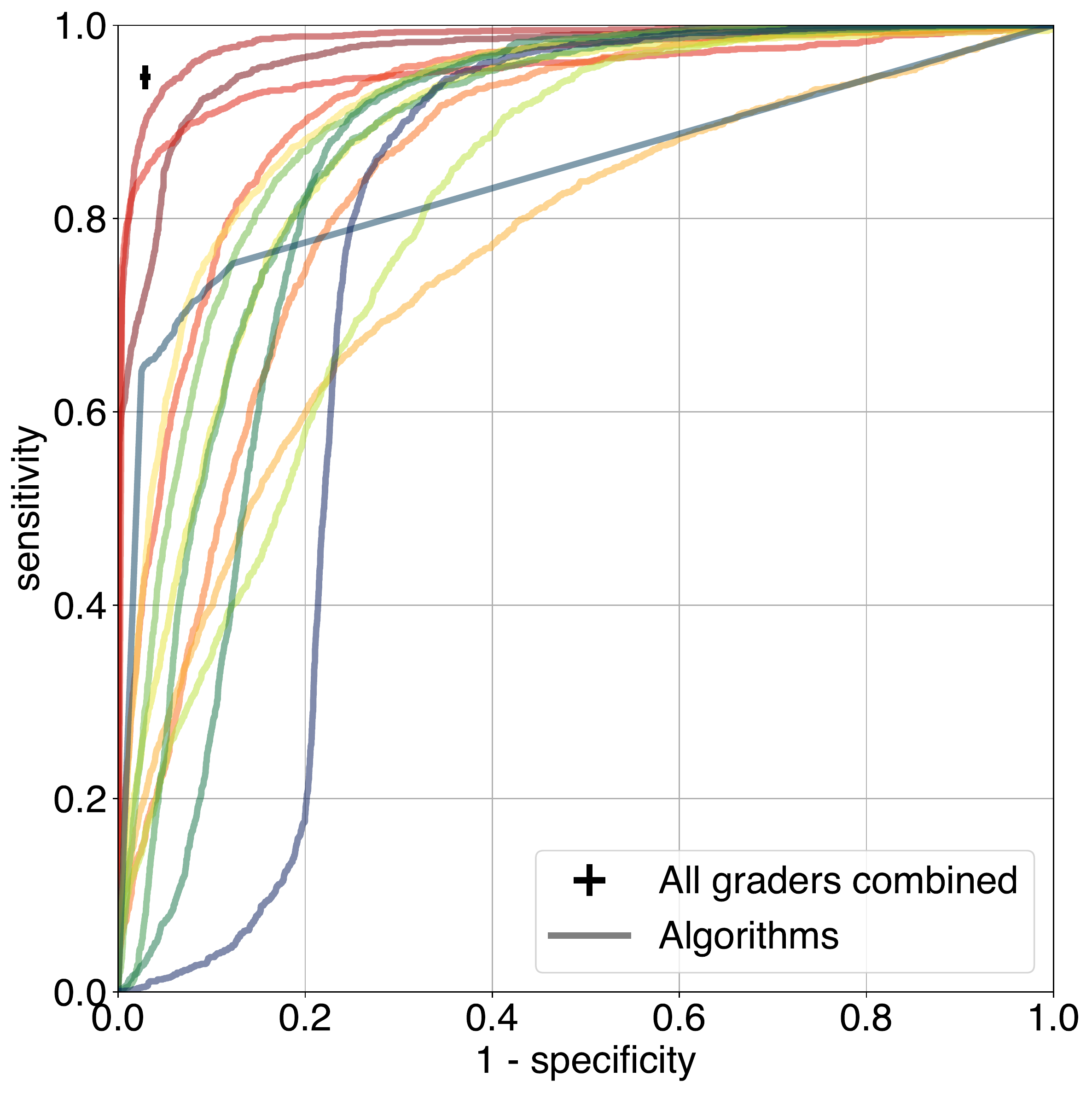}
    \caption{}
    \label{fig:roc_robustness}
\end{subfigure}

\caption{ROC curves for both challenge tasks. The sensitivity and specificity of all human graders on the AIROGS test set combined is indicated with black lines. Respectively, the width and height of the black horizontal and vertical line are 95\% CIs. In (a), the partial ROC curve (90\%-100\% specificity) for screening is shown, with 1,602 positive (RG) and 8,134 negative (NRG) images from the AIROGS test set. In (b), the ROC curve for robustness is shown with 1,554 positive (ungradable) and 9,736 negative (gradable) images from the AIROGS test set.}
\label{fig:ensembles}
\end{figure*}

\begin{figure}[!t]
\centering
\includegraphics[width=.35\linewidth]{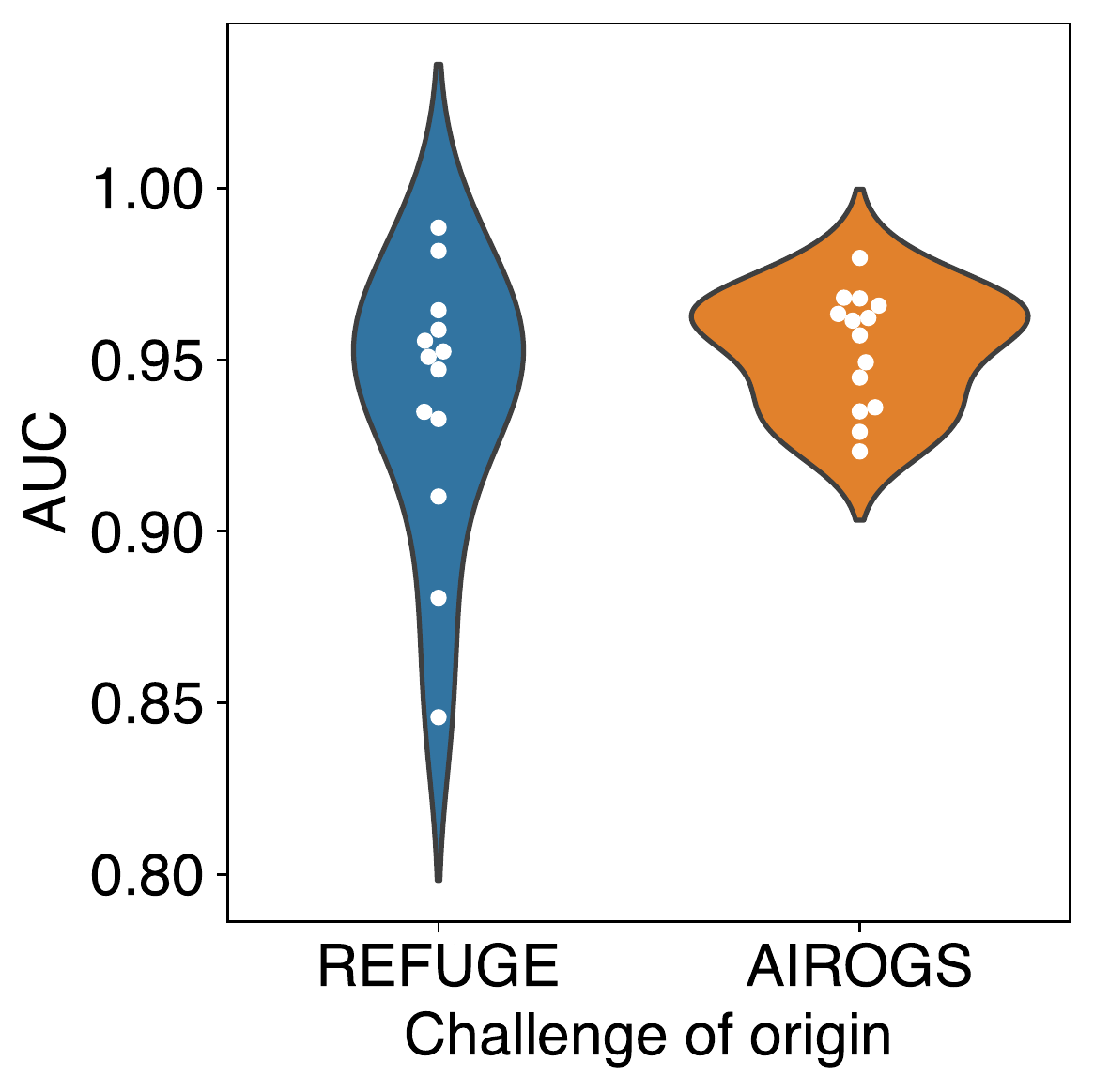}
\caption{Comparison of the AIROGS and REFUGE algorithms, tested on the REFUGE test set, visualized as violin and swarm plots. The final algorithms that were developed for the REFUGE challenge itself and for the AIROGS challenge are shown on the left and right, respectively. The AIROGS algorithms were only trained on the AIROGS train set and were not retrained with the REFUGE dataset.}
\label{fig:refuge_violin}
\end{figure}

\begin{figure}[!t]
\centering
\begin{subfigure}{0.33\linewidth}
    \includegraphics[width=\linewidth]{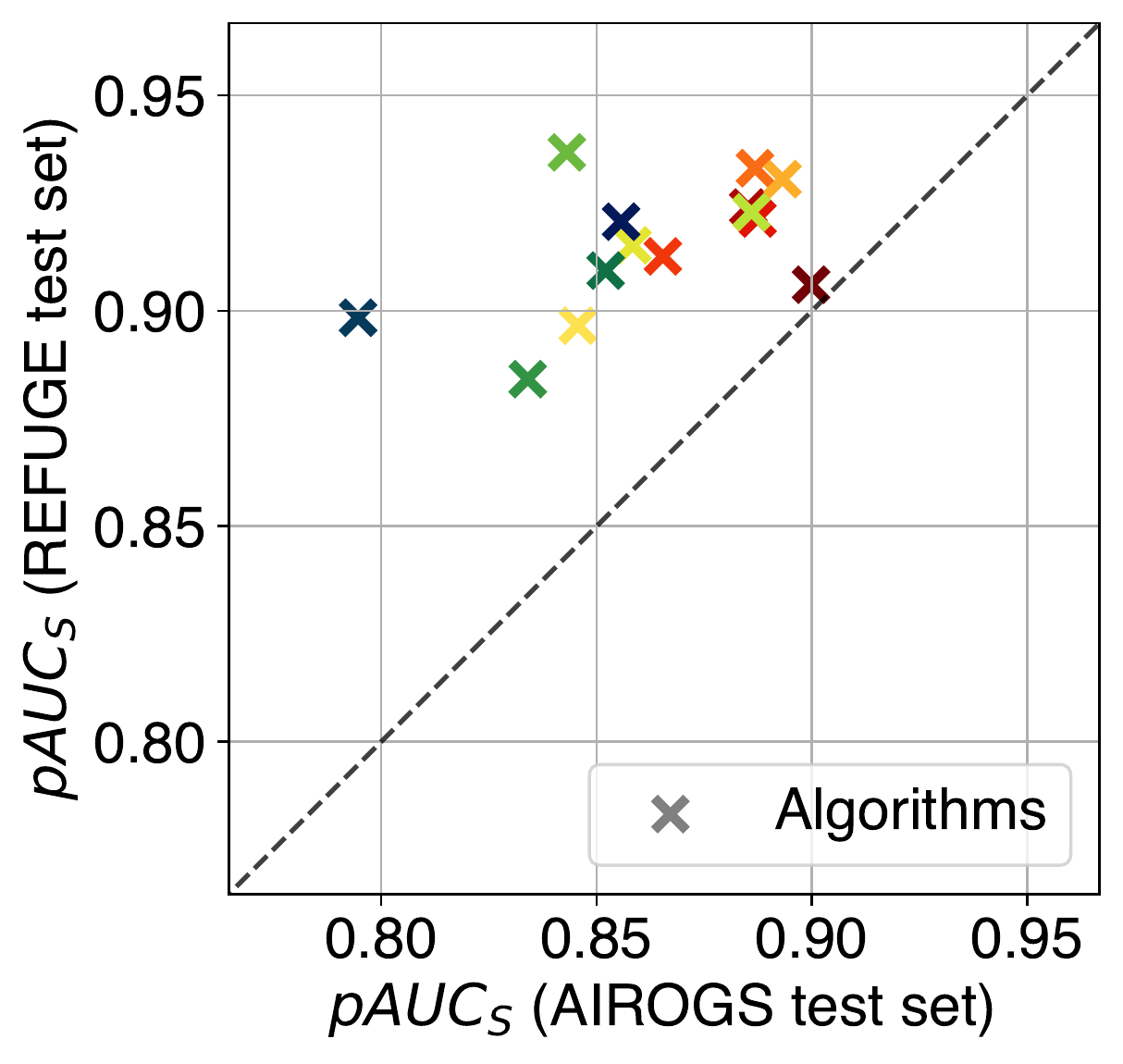}
    \caption{}
    \label{fig:refuge_alpha}
\end{subfigure}
\hspace{1cm}
\begin{subfigure}{0.33\linewidth}
    \includegraphics[width=\linewidth]{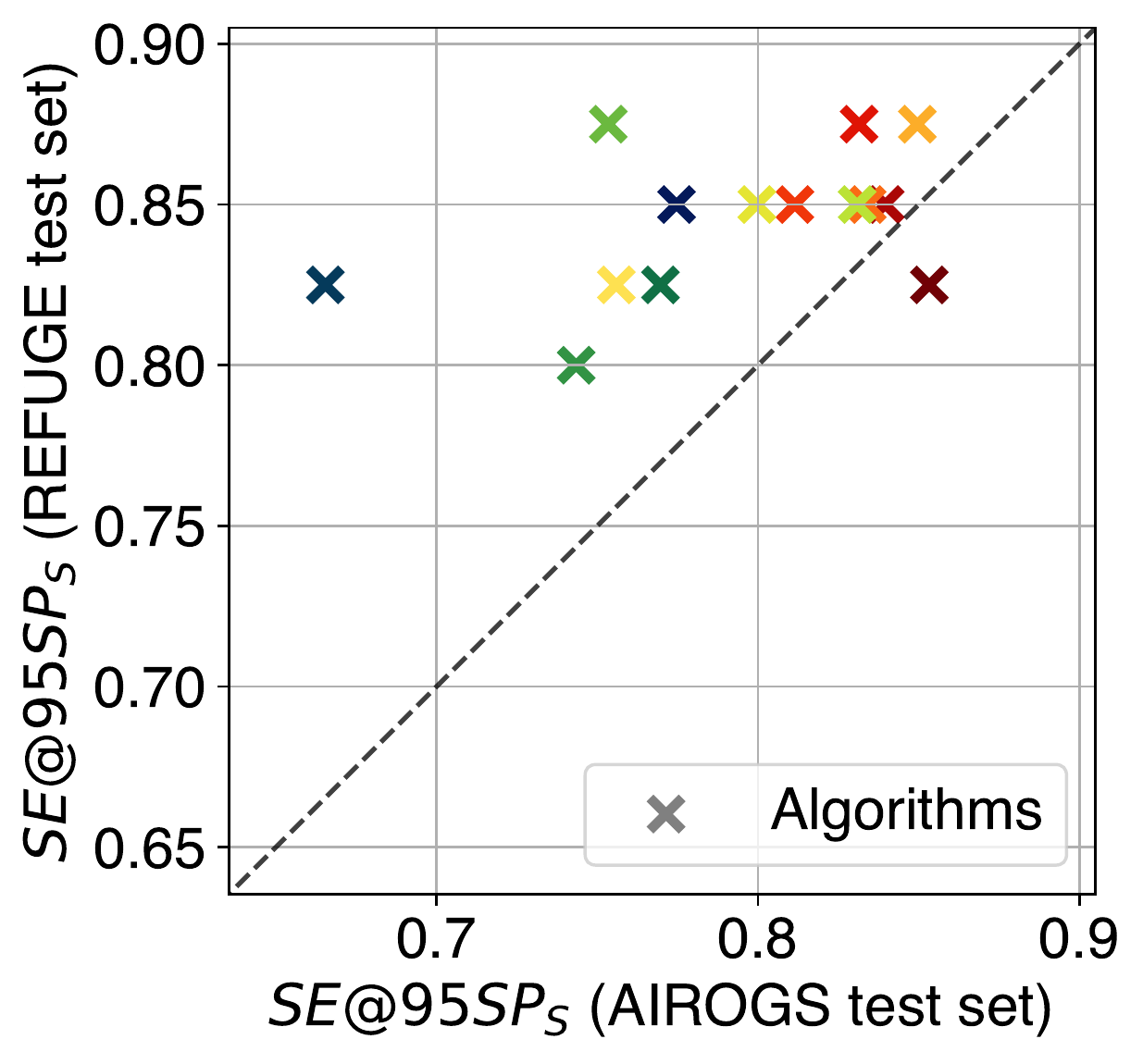}
    \caption{}
    \label{fig:refuge_beta}
\end{subfigure}
\caption{Performance of the participating AIROGS algorithms on the REFUGE dataset, compared to their performance on the AIROGS dataset. Both screening metrics (\protect\subref{fig:refuge_alpha}) $pAUC_S$ and (\protect\subref{fig:refuge_beta}) $SE@95SP_S$ are shown.}
\label{fig:external_refuge}
\end{figure}

\begin{figure}[!t]
\centering
\begin{subfigure}{0.33\linewidth}
    \includegraphics[width=\linewidth]{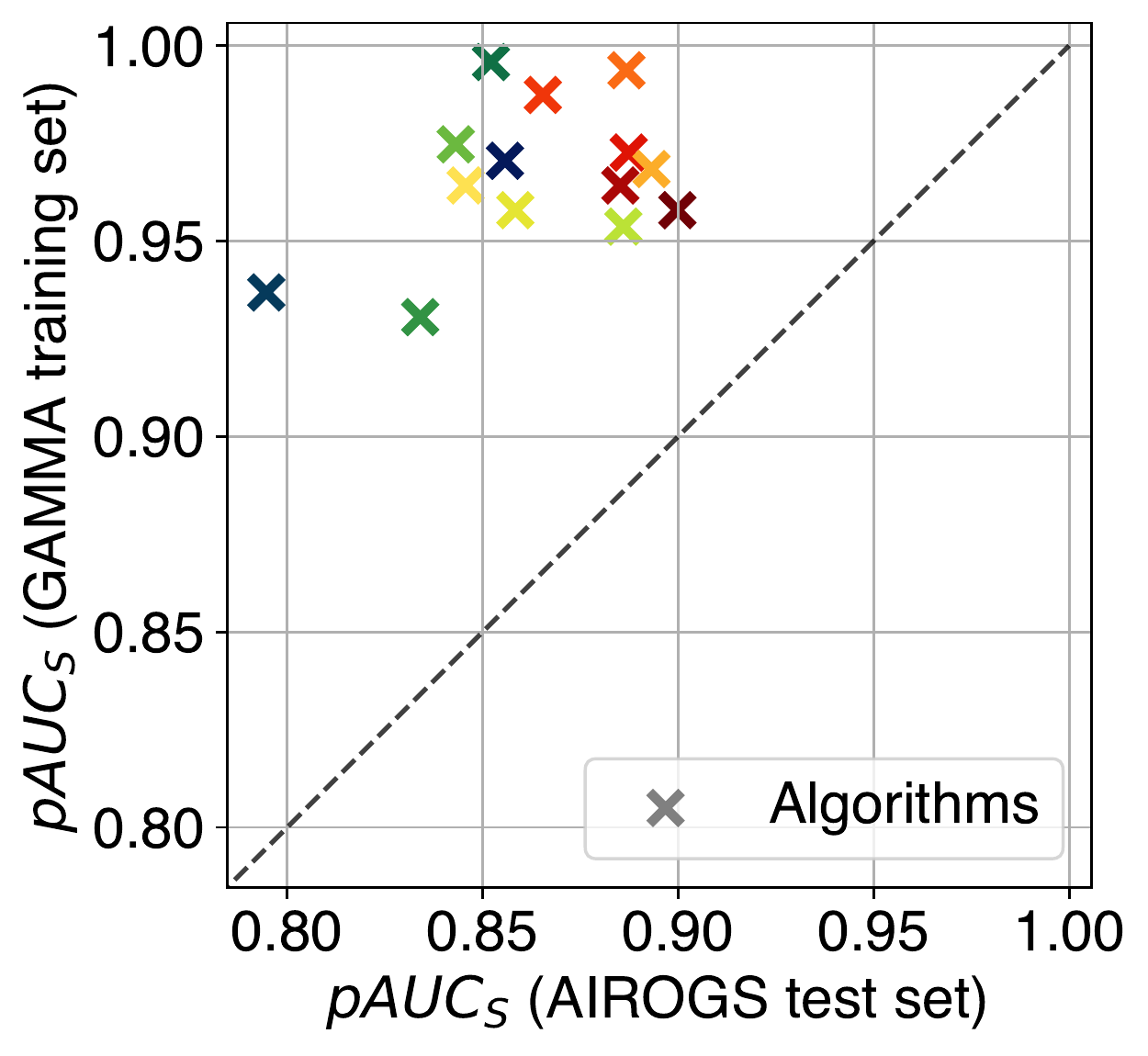}
    \caption{}
    \label{fig:gamma_alpha}
\end{subfigure}
\hspace{1cm}
\begin{subfigure}{0.33\linewidth}
    \includegraphics[width=\linewidth]{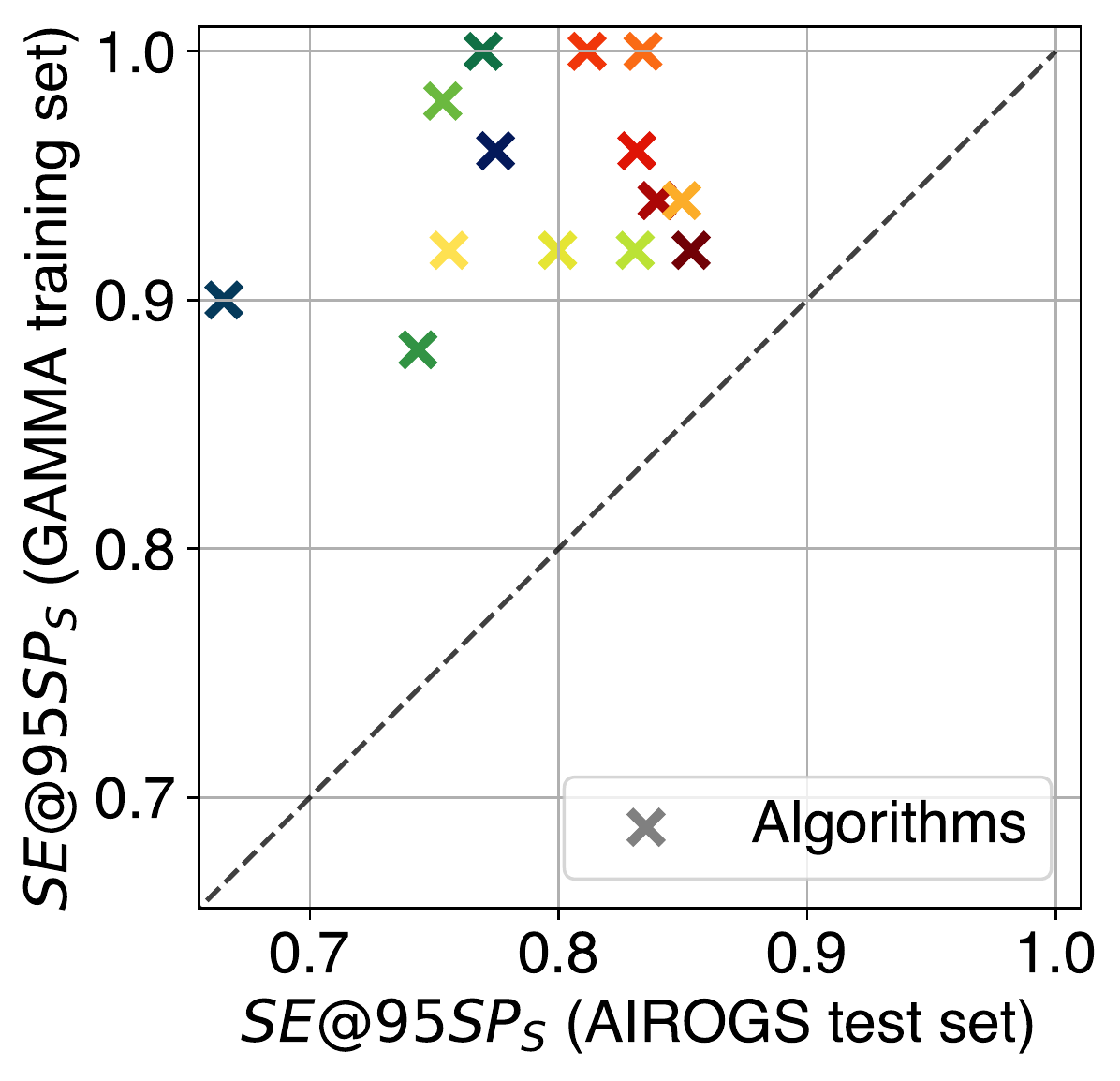}
    \caption{}
    \label{fig:gamma_beta}
\end{subfigure}
\caption{Performance of the participating AIROGS algorithms on the GAMMA dataset, compared to their performance on the AIROGS dataset. Both screening metrics (\protect\subref{fig:gamma_alpha}) $pAUC_S$ and (\protect\subref{fig:gamma_beta}) $SE@95SP_S$ are shown.}
\label{fig:external_gamma}
\end{figure}

\begin{figure}[!t]
\centering
\begin{subfigure}{0.33\linewidth}
    \includegraphics[width=\linewidth]{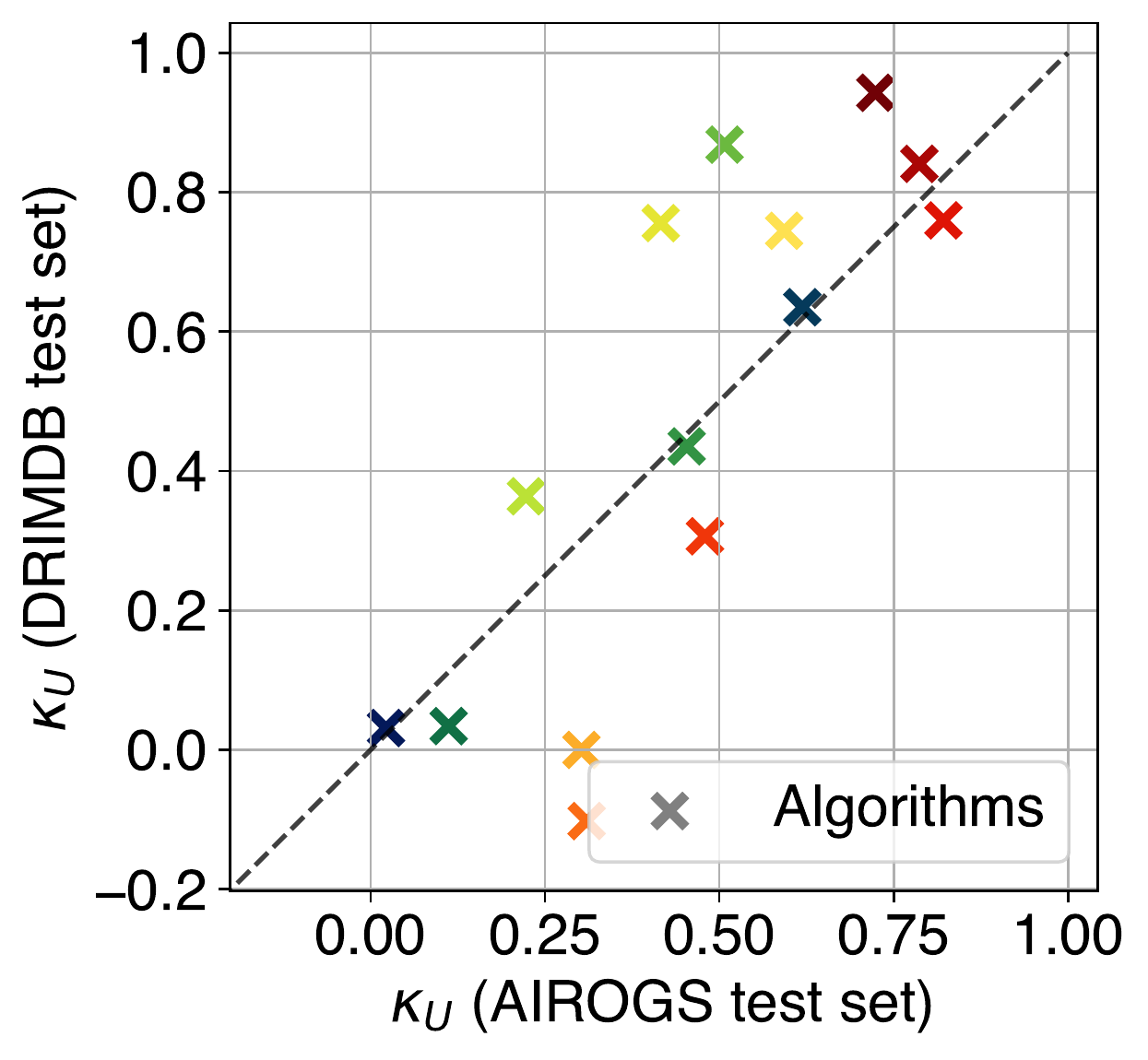}
    \caption{}
    \label{fig:drimdb_gamma}
\end{subfigure}
\hspace{1cm}
\begin{subfigure}{0.33\linewidth}
    \includegraphics[width=\linewidth]{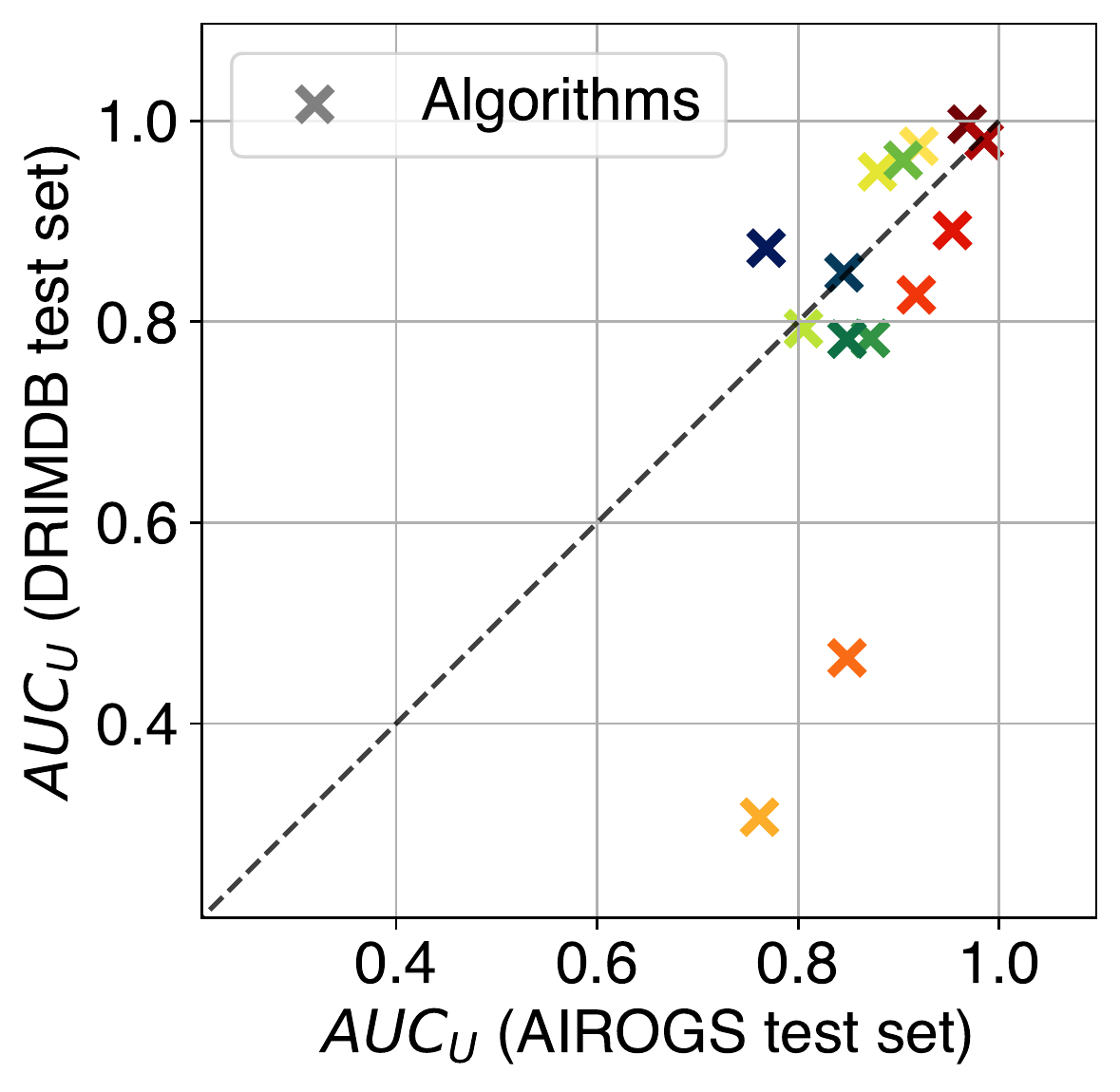}
    \caption{}
    \label{fig:drimdb_delta}
\end{subfigure}
\caption{Performance of the participating AIROGS algorithms on the DRIMDB dataset, compared to their performance on the AIROGS dataset. Both robustness metrics (\protect\subref{fig:drimdb_gamma}) $\kappa_U$ and (\protect\subref{fig:drimdb_delta}) $AUC_U$ are shown.}
\label{fig:external_drimdb}
\end{figure}

This section presents the glaucoma screening performance and robustness of the fourteen participating teams. The final rankings and mean positions of the teams are shown in the first plot of Fig. \ref{fig:all_metrics}. Four teams shared a rank with another team, since their mean positions were exactly equal, causing there to be two teams for each of the ranks \#2 and \#11.

\subsection{Glaucoma screening performance}


The glaucoma screening performance of the participating teams is summarized in Fig. \ref{fig:all_metrics}, showing $pAUC_S$ and $SE@95SP_S$ in the second and third plot, respectively. The highest scores for $pAUC_S$ and $SE@95SP_S$ were 0.90 (95\% CI: 0.89 -- 0.91) and 0.85 (95\% CI: 0.83 -- 0.87), respectively. These scores were both achieved by team \textit{PUMCH-eye}.

Fig. \ref{fig:ensembles_alpha} and Fig. \ref{fig:ensembles_beta} show the $pAUC_S$ and $SE@95SP_S$ for the ensembles when averaging the RG likelihood output $O_1$ of the best $M$ participants in terms of the relevant metric. An optimal $pAUC_S$ of 0.91 (95\% CI: 0.90 -- 0.92) was achieved at $M=3$. At $M=2$, an optimal value for $SE@95SP_S$ was reached, which was 0.87 (95\% CI: 0.85 -- 0.89).

Fig. \ref{fig:roc_screening} shows the partial ROC curves between 90\% and 100\% specificity for all participants. The plot also presents the sensitivity and specificity of the human graders with a 95\% CI. These were 0.86 (95\% CI: 0.84 -- 0.87) and 0.94 (95\% CI: 0.94 -- 0.95), respectively.


In Fig. \ref{fig:refuge_violin}, we compare the performance on the REFUGE test set of the final AIROGS algorithms, which were trained on the AIROGS train set, to the performance of the algorithms that were submitted to the REFUGE challenge, which were trained on the REFUGE train set. The top three participants of the REFUGE algorithms achieved AUCs of 0.99, 0.98 and 0.96. For the AIROGS algorithms, the best three AUCs were 0.98, 0.97 and 0.97. The mean $\pm$ std. dev. AUC of all REFUGE and AIROGS algorithms were 0.94 $\pm$ 0.04 and 0.95 $\pm$ 0.02, respectively. Fig. \ref{fig:external_refuge} presents the relation between the two glaucoma screening performance metrics of all participating AIROGS algorithms on the AIROGS test set and that performance on the REFUGE test set. For both metrics, almost all AIROGS algorithms (except for team \textit{PUMCH-eye} for $SE@95SP_S$) scored higher on REFUGE than on AIROGS. Of all AIROGS participants, the best $pAUC_S$ and $SE@95SP_S$ on REFUGE were 0.94 and 0.88, respectively.

In Fig. \ref{fig:external_gamma}, the relation between the glaucoma performance of all participating AIROGS algorithms on the AIROGS test set and that performance on GAMMA is shown. For both screening metrics, 
all AIROGS algorithms scored higher on GAMMA than on AIROGS. Of all AIROGS participants, the best $pAUC_S$ and $SE@95SP_S$ on GAMMA were 1.0 and 1.0, respectively.


\subsection{Robustness}

The robustness metrics of the participating teams are summarized in Fig. \ref{fig:all_metrics}, showing $\kappa_U$ and $AUC_U$ in the fourth and fifth plot, respectively. The highest scores for $\kappa_U$ and $AUC_U$ were 0.82 (95\% CI: 0.80 -- 0.84) and 0.99 (95\% CI: 0.98 -- 0.99), respectively. These scores were achieved by team \textit{Temirgali} and \textit{RWTH-CuP}, respectively.

Fig. \ref{fig:ensembles_gamma} and Fig. \ref{fig:ensembles_delta} show the $\kappa_U$ and $AUC_U$ for the ensembles when averaging output $O_3$ and output $O_4$, respectively, of the $M$ best algorithms in terms of these respective metrics. An optimal $\kappa_U$ of 0.85 (95\% CI: 0.84 -- 0.86) was achieved at $M=6$. Also at $M=6$, an optimal value for $AUC_U$ was reached, which was 0.99 (95\% CI: 0.99 -- 0.99).

In Fig. \ref{fig:roc_robustness}, ROC curves for robustness are shown for all participants. The plot also presents the sensitivity and specificity for separating ungradable from gradable images of the human graders with a 95\% CI. These were 0.95 (95\% CI: 0.94 -- 0.96) and 0.97 (95\% CI: 0.97 -- 0.97), respectively.

The results on the external DRIMDB dataset are shown in Fig. \ref{fig:external_drimdb}, indicating the relation between the ungradability metrics $\kappa_U$ and $AUC_U$ of all participating AIROGS algorithms on DRIMDB and those metrics on AIROGS. Of all AIROGS participants, the best $\kappa_U$ and $AUC_U$ on DRIMDB were 0.94 and 1.0, respectively.



\section{Discussion}
AI models have been shown to be effective at detecting glaucoma in CFPs, but most studies lack evidence of robustness to real-world scenarios in which unexpected OOD data can be presented due to various causes. To this end, we relied on the community to develop robust AI solutions for glaucoma screening based on the largest multi-center real-world CFP dataset with glaucoma labels. We organized the AIROGS challenge around this dataset, ensuring the resulting algorithms are reusable in a cloud-based environment. We applied these algorithms to ungradable data, while the participants could only train on gradable data to ensure robustness to any kind of ungradable data, and to other publicly available datasets to assess their generalization.

\subsection{Overall findings}
The team with the highest $SE@95SP_S$ scored expert-level screening performance on the AIROGS test set with a sensitivity of 0.85 (95\% CI: 0.83 – 0.87) at 95\% specificity, similar to the sensitivity of 0.86 (95\% CI: 0.84 – 0.87) at a specificity of 0.94 (95\% CI: 0.94 – 0.95) of human graders. The highest $pAUC_S$ that was achieved by any of the teams was 0.90 (95\% CI: 0.89 – 0.91). Ensembling the different participating methods improved the screening performance even further, to 0.91 (95\% CI: 0.90 – 0.92) and 0.87 (95\% CI: 0.85 – 0.89) for $pAUC_S$ and $SE@95SP_S$, respectively. Seven out of fourteen teams exceeded the minimum performance of 80\% sensitivity and 95\% specificity that was required by human graders who were periodically monitored during the grading process. This shows these models can provide a similar performance to human graders for glaucoma screening, suggesting that AI can potentially play a role in an automated screening process.

We also evaluated the screening performance of the algorithms on two external test sets. Even though the algorithms were trained on AIROGS data, they achieved very high performances on the two external test sets, showing reproducible results in different sets and populations. On average, the participating AIROGS algorithms scored slightly higher on the REFUGE dataset than the REFUGE participants. We found that the participating algorithms scored substantially higher on these external datasets than on the AIROGS test set, indicating the value of a challenging real-world dataset. This strong generalization of the developed solutions also shows the potential of models trained on our dataset to be successfully implemented in screening programs with limited to no loss of performance.

The robustness to ungradable data in the AIROGS test set was evaluated for each team using the metrics $\kappa_U$ and $AUC_U$. The teams that performed the best in terms of these metrics achieved 0.82 (95\% CI: 0.80 – 0.84) and 0.99 (95\% CI: 0.98 – 0.99) for $\kappa_U$ and $AUC_U$, respectively. Human experts did reach a higher $\kappa_U$ of 0.85 (95\% CI: 0.84-0.86) for this task. Moreover, they achieved a sensitivity of 0.95 (95\% CI: 0.94 – 0.96) and a specificity of 0.97 (95\% CI: 0.97 – 0.97) for detecting ungradable cases, while the team with the best $AUC_U$ achieved a lower sensitivity at 97\% specificity of 0.90 (95\% CI: 0.88-0.92). Although the teams achieved relatively high performances, they still achieved lower performance at the robustness task than human experts. This shows this task was especially challenging, possibly because the participating teams could not use ungradable development data or because their robustness approaches focused on specific forms of ungradability.

We also assessed robustness on the external DRIMDB dataset. The best scoring team on this dataset scored very high performances; they achieved 0.94 and 1.0 for $\kappa_U$ and $AUC_U$, respectively. These two metrics were lower on the AIROGS dataset for that team. This also indicates very strong generalization to other datasets for the robustness task. The high ungradability detection performance also indicates robustness to other diseases in the image, as diabetic retinopathy was prevalent in the gradable subset of DRIMDB and the best algorithms did not classify these diseases as ungradable.

A large difference in performance between participating teams can be observed, both for the screening and the robustness task. Therefore, we think it is important to identify which methodological choices were made predominantly by top performing teams.
One of the most notable differences between the top three participants and the rest was the use of transformers. Outside of the top three, only the latest placed team used a transformer. Moreover, all best three participants manually labelled ODs for training either a segmentation or detection model to crop around the OD during pre-processing. Even though this was also done by two other teams, this seems like an effective strategy to achieve higher screening performance. A likely reason for the effectiveness of this approach is that most glaucoma-related imaging features can be found on or around the OD. This shows how a priori medical knowledge could still be of value even when large amount of data is available. A less important factor appears to be the number of manually labeled ODs. A possible reason for this could be that the OD detection or segmentation network is not required to be extremely accurate as combining a rough localization of the OD with a large enough padding margin could also suffice to crop the image during pre-processing.


Since the development set only consisted of images that were labeled gradable (either RG or NRG) and the use of external fundus data was prohibited, all teams came up with an uncertainty or OOD detection method based on the gradable data for the robustness task. The ungradability methods of the top three participants in terms of mean position, $\kappa_U$ and $AUC_U$, all revolved around the confidence of a neural network that localized the OD. Of the other participants, only the ninth placed team had such an approach. Apart from these methods based on OD detection, only team \textit{UPF+AIML} implemented a different robustness technique that was also based on domain-knowledge. This raises the impression that solutions based on domain knowledge are more effective for robustness than more general OOD detection solutions. However, it still needs to be evaluated if such approaches are robust for other general tasks (not glaucoma screening) or other sources of OOD data.

For calculating the $\kappa_U$ metric, the participants were required to output a binary decision on ungradability. A popular approach, especially among the top participants, was to manually identify relatively low quality images in the development set and base a threshold for this binary output on that subset. This technique was employed by the best three, fifth, tenth and twelfth team in terms of $\kappa_U$. This indicates that this could be a successful approach, although not in general as the accuracy of this binary value is also highly dependent on the quality of the scalar output for ungradability $O_4$ that is being thresholded. We found the difference between the ranking in terms of $\kappa_U$ and $AUC_U$ of one team in particular stood out. Team \textit{YC} ranked only eleventh for $AUC_U$ (which depended on the scalar output $O_4$), but ranked fourth in terms of $\kappa_U$ (which depended on the binary output $O_3$), indicating the approach they used for thresholding their scalar value was highly effective. The difference between their $AUC_U$ and $\kappa_U$ was 7, while the next biggest value of this difference was only 3. Team \textit{YC} indeed came up with a relatively sophisticated method for binarizing $O_4$ compared to others, based on a Wilcoxon one-sample test to statistically test whether the mean of the predicted probabilties from a Monte-Carlo drop-out approach was 0.5 or not. 

\subsection{Strengths and limitations}


The dataset presented in this paper substantially exceeds what was publicly available before in terms of number of images and patients. The dataset is also highly diverse because of the large number of different sites, cameras that were used and ethnicities. The quality of the labels was controlled by the initial and periodical evaluation of human graders, the fact that each image was independently labeled twice by two trained graders and, in case of disagreement, by a highly experienced reader. The participants submitted their solutions as containerized algorithms, allowing reproducibility, facilitating inference on other data, and preventing manual manipulation of the test set.

One of the rules of the AIROGS challenge was the prohibition of the use of external fundus data for development. A limitation of this work is the fact that we cannot be sure if any of the teams used such data in their development process. 
A possible approach to prevent this and making the process fairer is to have participants submit a containerized algorithm for training, which would be trained by the challenge organizers with private challenge training data. Nevertheless, with such an approach it would still be challenging and time-consuming for the challenge organizers to verify if the training containers do not contain any weights pre-trained on other data.

The dataset used for the challenge is diverse, but improvements could still be made in that respect. All screening sites were based across the United States of America, raising the question whether a more generalizable model could be obtained with data from across the world. On the other hand, we showed that many algorithms trained on the AIROGS dataset performed at least as well on three external test sets, of which two originated from China and one from Turkey, as on our internal test set.

Not all research groups working in the field of retinal image analysis participated in this challenge and many teams that joined the challenge did not submit a solution to the \textit{Final Test Phase}. Possible reasons for this include that many teams saw their results did not match to ones already present on the leaderboard, that the barrier for some teams was too high to get a solution wrapped in a Docker container, or that they were not able to finish in time. Therefore we would like to stress the challenge is still open and we are curious to see if the community can make further improvements. After all, especially for the robustness task, there seems to be room for improvement, given the gap with the human grader performance.

\subsection{Future directions}
Based on the solutions that were presented by the teams, we think it would be valuable to combine methodologies from different participants and to work further on their ideas. For example, as we mentioned before, team \textit{YC} apparently had a highly effective method for thresholding their ungradability scores as their $\kappa_U$ was very high compared to their $AUC_U$. A possible future direction would be to combine methods of high performance in terms of $AUC_U$ with the binarization technique from team \textit{YC}. Moreover, we observed that algorithms which scored high in terms of robustness, used domain-knowledge for this aspect of the challenge. Possible future directions could be to explore other ways to incorporate domain-knowledge into an ungradability method.
This observation also leads to the question whether there are more fields in medical image analysis in which domain-knowledge can be leveraged for uncertainty estimation and OOD detection.

Next to a decision on RG and NRG presence, the graders were asked to provide which clinical, glaucomatous features were present in the eyes they classified as RG, as listed in Section \ref{sec:datasets:challenge_dataset} and further described by \cite{Lemi22}. This information was not yet included in the dataset release for this challenge, as it fell outside the scope of this challenge. Future solutions and challenges could be developed with this information, possibly resulting in more explainable algorithms.

This challenge only focused on classification based on a single CFP. It may be interesting to explore the effect on screening performance and robustness of including various types of metadata in our dataset, which we have available but have not been published yet. This metadata, although missing for some images, includes the camera type, age and anonymous patient identification (which can be used to link two eyes to a single patient).

\section{Conclusions}
We presented the results of community-acquired algorithms tested on real-world data for robust glaucoma screening from CFP. The best algorithms performed similarly in terms of screening to the carefully trained and selected human graders, and were shown to be effective at flagging images that could not be graded. Methodological choices predominantly made by the best teams included, for the screening task, the use of vision transformers and the incorporation of optic disc detection models in pre-processing and, for the robustness task, out-of-distribution detection approaches based on domain-knowledge. We hope the unprecedented size and real-world nature of the dataset we released and the algorithms that were developed using this dataset will help towards implementing robust AI for glaucoma screening.

\section*{Acknowledgement}
\footnotesize{Coen de Vente and Clara I. S\'{a}nchez are with the Quantitative Healthcare Analysis (QurAI) Group, Informatics Institute, Universiteit van Amsterdam, Amsterdam, Noord-Holland, Netherlands and the Department of Biomedical Engineering and Physics, Amsterdam UMC Locatie AMC, Amsterdam, Noord-Holland, Netherlands (e-mail: research@coendevente.com).

Coen de Vente and Bram van Ginneken are with the Diagnostic Image Analysis Group (DIAG), Department of Radiology and Nuclear Medicine, Radboudumc, Nijmegen, Gelderland, Netherlands.

Koenraad A. Vermeer and Hans G. Lemij are with the Rotterdam Ophthalmic Institute, Rotterdam Eye Hospital, Rotterdam, Netherlands.

Nicolas Jaccard is with Project Orbis International Inc., New York, United States.

He Wang is with the Peking Union Medical College Hospital, Beijing, China and with the Xuanwu Hospital Capital Medical University, Beijing, 100053, China.

Hongyi Sun is with the Tsinghua University, Beijing, China.

Firas Khader and Daniel Truhn are with the Department of Diagnostic and Interventional Radiology, University Hospital Aachen, Aachen, Germany.

Temirgali Aimyshev and Yerkebulan Zhanibekuly are with CMC Technologies LLP, Nur-Sultan, Kazakhstan.

Tien-Dung Le is with KBC, Belgium.

Adrian Galdran and Miguel \'Angel Gonz\'alez-Ballester are with Universitat Pompeu Fabra, Barcelona, Spain.

Adrian Galdran and Gustavo Carneiro are with the Australian Institute for Machine Learning AIML, University of Adelaide, Australia.

\changed{Miguel \'Angel Gonz\'alez-Ballester is also with ICREA, Barcelona, Spain.}

\changed{Gustavo Carneiro is also with the Centre for Vision, Speech and Signal Processing, University of Surrey, United Kingdom.}

Devika R G is with the College of Engineering, Trivandrum, India.

Hrishikesh P S and Densen Puthussery are with Founding Minds Software, India.

Hong Liu and Zekang Yang are with the Institute of Computing Technology, Chinese Academy of Sciences.

Satoshi Kondo is with the Muroran Institute of Technology, Japan.

Satoshi Kasai is with the Niigata University of Health and Welfare, Japan.

Edward Wang and Ashritha Durvasula are with the Schulich School of Medicine and Dentistry, University of Western Ontario, London, Canada.

J\'{o}nathan Heras is with the Department of Mathematics and Computer Science, University of La Rioja, Spain.

Miguel \'{A}ngel Zapata is with Hospital Vall Hebron, Passeig Roser 126, Sant Cugat del Vallès, 08195 Barcelona, Spain and UPRetina, Barcelona, Spain.

Teresa Ara\'{u}jo, Guilherme Aresta and Hrvoje Bogunovi\'{c} are with the Christian Doppler Laboratory for Artificial Intelligence in Retina, Department of Ophthalmology and Optometry, Medical University of Vienna, Vienna, Austria.

Mustafa Arikan is with the Institute of Ophthalmology, University College London, London, United Kingdom.

Yeong Chan Lee is with the Research Institute for Future Medicine, Samsung Medical Center, Seoul, Republic of Korea.

Hyun Bin Cho and Yoon Ho Choi are with the Department of Digital Health, Samsung Advanced Institute for Health Sciences \& Technology (SAIHST), Sungkyunkwan University, Samsung Medical Center, Seoul, Republic of Korea.

Yoon Ho Choi is also with the Department of Artificial Intelligence and Informatics, Mayo Clinic, United States of America, Florida, Jacksonville.

Abdul Qayyum is with the Department of Biomedical Engineering, King’s College London, UK.

Imran Razzak is with the University of New South Wales, Sydney, Australia.}

\bibliographystyle{unsrt}  
\bibliography{bib/newrefs}

\begin{thebibliography}{10}

\bibitem{Tham14}
Yih-Chung Tham, Xiang Li, Tien~Y Wong, Harry~A Quigley, Tin Aung, and Ching-Yu
  Cheng.
\newblock Global prevalence of glaucoma and projections of glaucoma burden
  through 2040: a systematic review and meta-analysis.
\newblock {\em Ophthalmology}, 121(11):2081--2090, 2014.

\bibitem{Mokh16}
Palwasha Mokhles, Jan~SAG Schouten, Henny~JM Beckers, Augusto Azuara-Blanco,
  Anja Tuulonen, and Carroll~AB Webers.
\newblock A systematic review of end-of-life visual impairment in open-angle
  glaucoma: an epidemiological autopsy.
\newblock {\em Journal of Glaucoma}, 25(7):623--628, 2016.

\bibitem{Erne13}
Paul~JG Ernest, Michiel~JWM Busch, Carroll~AB Webers, Henny~JM Beckers, Fred
  Hendrikse, Martin~H Prins, and Jan~SAG Schouten.
\newblock Prevalence of end-of-life visual impairment in patients followed for
  glaucoma.
\newblock {\em Acta Ophthalmologica}, 91(8):738--743, 2013.

\bibitem{Pete13}
Dorothea Peters, Boel Bengtsson, and Anders Heijl.
\newblock Lifetime risk of blindness in open-angle glaucoma.
\newblock {\em American Journal of Ophthalmology}, 156(4):724--730, 2013.

\bibitem{Wein14}
Robert~N Weinreb, Tin Aung, and Felipe~A Medeiros.
\newblock The pathophysiology and treatment of glaucoma: a review.
\newblock {\em JAMA}, 311(18):1901--1911, 2014.

\bibitem{Chen15}
Xiangyu Chen, Yanwu Xu, Fengshou Yin, Zhuo Zhang, Damon Wing~Kee Wong, Tien~Yin
  Wong, and Jiang Liu.
\newblock Multiple ocular diseases detection based on joint sparse multi-task
  learning.
\newblock In {\em 2015 37th Annual International Conference of the IEEE
  Engineering in Medicine and Biology Society (EMBC)}, pages 5260--5263. IEEE,
  2015.

\bibitem{Ting17}
Daniel Shu~Wei Ting, Carol Yim-Lui Cheung, Gilbert Lim, Gavin Siew~Wei Tan,
  Nguyen~D Quang, Alfred Gan, Haslina Hamzah, Renata Garcia-Franco, Ian~Yew
  San~Yeo, Shu~Yen Lee, et~al.
\newblock Development and validation of a deep learning system for diabetic
  retinopathy and related eye diseases using retinal images from multiethnic
  populations with diabetes.
\newblock {\em JAMA}, 318(22):2211--2223, 2017.

\bibitem{Li18}
Zhixi Li, Yifan He, Stuart Keel, Wei Meng, Robert~T Chang, and Mingguang He.
\newblock Efficacy of a deep learning system for detecting glaucomatous optic
  neuropathy based on color fundus photographs.
\newblock {\em Ophthalmology}, 125(8):1199--1206, 2018.

\bibitem{Phen19}
Sonia Phene, R~Carter Dunn, Naama Hammel, Yun Liu, Jonathan Krause, Naho
  Kitade, Mike Schaekermann, Rory Sayres, Derek~J Wu, Ashish Bora, et~al.
\newblock Deep learning and glaucoma specialists: the relative importance of
  optic disc features to predict glaucoma referral in fundus photographs.
\newblock {\em Ophthalmology}, 126(12):1627--1639, 2019.

\bibitem{Roge19}
Thomas~W Rogers, Nicolas Jaccard, Francis Carbonaro, Hans~G Lemij, Koenraad~A
  Vermeer, Nicolaas~J Reus, and Sameer Trikha.
\newblock Evaluation of an ai system for the automated detection of glaucoma
  from stereoscopic optic disc photographs: the european optic disc assessment
  study.
\newblock {\em Eye}, 33(11):1791--1797, 2019.

\bibitem{Beed20}
Emma Beede, Elizabeth Baylor, Fred Hersch, Anna Iurchenko, Lauren Wilcox,
  Paisan Ruamviboonsuk, and Laura~M Vardoulakis.
\newblock A human-centered evaluation of a deep learning system deployed in
  clinics for the detection of diabetic retinopathy.
\newblock In {\em Proceedings of the 2020 CHI Conference on Human Factors in
  Computing Systems}, pages 1--12, 2020.

\bibitem{Zhan10}
Zhuo Zhang, Feng~Shou Yin, Jiang Liu, Wing~Kee Wong, Ngan~Meng Tan, Beng~Hai
  Lee, Jun Cheng, and Tien~Yin Wong.
\newblock Origa-light: An online retinal fundus image database for glaucoma
  analysis and research.
\newblock In {\em 2010 Annual International Conference of the IEEE Engineering
  in Medicine and Biology}, pages 3065--3068. IEEE, 2010.

\bibitem{Fume11}
Francisco Fumero, Silvia Alay{\'o}n, Jos{\'e}~L Sanchez, Jose Sigut, and
  M~Gonzalez-Hernandez.
\newblock Rim-one: An open retinal image database for optic nerve evaluation.
\newblock In {\em 24th International Symposium on Computer-Based Medical
  Systems (CBMS)}, pages 1--6. IEEE, 2011.

\bibitem{Odst13}
Jan Odstrcilik, Radim Kolar, Attila Budai, Joachim Hornegger, Jiri Jan, Jiri
  Gazarek, Tomas Kubena, Pavel Cernosek, Ondrej Svoboda, and Elli Angelopoulou.
\newblock Retinal vessel segmentation by improved matched filtering: evaluation
  on a new high-resolution fundus image database.
\newblock {\em IET Image Processing}, 7(4):373--383, 2013.

\bibitem{Siva15}
Jayanthi Sivaswamy, S~Krishnadas, Arunava Chakravarty, G~Joshi, A~Syed Tabish,
  et~al.
\newblock A comprehensive retinal image dataset for the assessment of glaucoma
  from the optic nerve head analysis.
\newblock {\em JSM Biomedical Imaging Data Papers}, 2(1):1004, 2015.

\bibitem{Holm17}
Sven Holm, Greg Russell, Vincent Nourrit, and Niall McLoughlin.
\newblock Dr hagis—a fundus image database for the automatic extraction of
  retinal surface vessels from diabetic patients.
\newblock {\em Journal of Medical Imaging}, 4(1):014503, 2017.

\bibitem{Orla18}
Jos{\'e}~Ignacio Orlando, Jo{\~a}o Barbosa~Breda, Karel~van Keer, Matthew~B
  Blaschko, Pablo~J Blanco, and Carlos~A Bulant.
\newblock Towards a glaucoma risk index based on simulated hemodynamics from
  fundus images.
\newblock In {\em International Conference on Medical Image Computing and
  Computer-Assisted Intervention}, pages 65--73. Springer, 2018.

\bibitem{Orla20}
Jos{\'e}~Ignacio Orlando, Huazhu Fu, Jo{\~a}o~Barbosa Breda, Karel van Keer,
  Deepti~R Bathula, Andr{\'e}s Diaz-Pinto, Ruogu Fang, Pheng-Ann Heng, Jeyoung
  Kim, JoonHo Lee, et~al.
\newblock {REFUGE} challenge: A unified framework for evaluating automated
  methods for glaucoma assessment from fundus photographs.
\newblock {\em Medical image analysis}, 59:101570, 2020.

\bibitem{Fang22}
Huihui Fang, Fei Li, Huazhu Fu, Xu~Sun, Xingxing Cao, Jaemin Son, Shuang Yu,
  Menglu Zhang, Chenglang Yuan, Cheng Bian, et~al.
\newblock Refuge2 challenge: Treasure for multi-domain learning in glaucoma
  assessment.
\newblock {\em arXiv preprint arXiv:2202.08994}, 2022.

\bibitem{Wu22}
Junde Wu, Huihui Fang, Fei Li, Huazhu Fu, Fengbin Lin, Jiongcheng Li, Lexing
  Huang, Qinji Yu, Sifan Song, Xingxing Xu, et~al.
\newblock Gamma challenge: glaucoma grading from multi-modality images.
\newblock {\em arXiv preprint arXiv:2202.06511}, 2022.

\bibitem{Cuad09}
Jorge Cuadros and George Bresnick.
\newblock Eyepacs: an adaptable telemedicine system for diabetic retinopathy
  screening.
\newblock {\em Journal of diabetes science and technology}, 3(3):509--516,
  2009.

\bibitem{Reus10}
Nicolaas~J Reus, Hans~G Lemij, David~F Garway-Heath, P~Juhani Airaksinen,
  Alfonso Anton, Alain~M Bron, Christoph Faschinger, G{\'a}bor Holl{\'o},
  Michele Iester, Jost~B Jonas, et~al.
\newblock Clinical assessment of stereoscopic optic disc photographs for
  glaucoma: the european optic disc assessment trial.
\newblock {\em Ophthalmology}, 117(4):717--723, 2010.

\bibitem{Sevi14}
Ugur Sevik, Cemal Kose, Tolga Berber, and Hidayet Erdol.
\newblock Identification of suitable fundus images using automated quality
  assessment methods.
\newblock {\em Journal of biomedical optics}, 19(4):046006, 2014.

\bibitem{Mccl89}
Donna~Katzman McClish.
\newblock Analyzing a portion of the roc curve.
\newblock {\em Medical decision making}, 9(3):190--195, 1989.

\bibitem{Vaah07}
Hanna Vaahtoranta-Lehtonen, Anja Tuulonen, Pasi Aronen, Harri Sintonen, Liisa
  Suoranta, Niina Kovanen, Miika Linna, Esa L{\"a}{\"a}r{\"a}, and Antti
  Malmivaara.
\newblock Cost effectiveness and cost utility of an organized screening
  programme for glaucoma.
\newblock {\em Acta Ophthalmologica Scandinavica}, 85(5):508--518, 2007.

\bibitem{Vrie12}
Margriet~M de~Vries, Remco Stoutenbeek, Rogier~PHM M{\"u}skens, and Nomdo~M
  Jansonius.
\newblock Glaucoma screening during regular optician visits: the feasibility
  and specificity of screening in real life.
\newblock {\em Acta ophthalmologica}, 90(2):115--121, 2012.

\bibitem{Rutt20}
Carolyn~M Rutter.
\newblock Bootstrap estimation of diagnostic accuracy with patient-clustered
  data.
\newblock {\em Academic radiology}, 7(6):413--419, 2000.

\bibitem{King14}
Diederik~P Kingma and Jimmy Ba.
\newblock Adam: A method for stochastic optimization.
\newblock {\em arXiv preprint arXiv:1412.6980}, 2014.

\bibitem{Losh17}
Ilya Loshchilov and Frank Hutter.
\newblock Decoupled weight decay regularization.
\newblock {\em arXiv preprint arXiv:1711.05101}, 2017.

\bibitem{Fore20}
Pierre Foret, Ariel Kleiner, Hossein Mobahi, and Behnam Neyshabur.
\newblock Sharpness-aware minimization for efficiently improving
  generalization.
\newblock {\em arXiv preprint arXiv:2010.01412}, 2020.

\bibitem{He22}
Kaiming He, Xinlei Chen, Saining Xie, Yanghao Li, Piotr Doll{\'a}r, and Ross
  Girshick.
\newblock Masked autoencoders are scalable vision learners.
\newblock In {\em Proceedings of the IEEE/CVF Conference on Computer Vision and
  Pattern Recognition}, pages 16000--16009, 2022.

\bibitem{Xie20}
Qizhe Xie, Minh-Thang Luong, Eduard Hovy, and Quoc~V Le.
\newblock Self-training with noisy student improves imagenet classification.
\newblock In {\em Proceedings of the IEEE/CVF conference on computer vision and
  pattern recognition}, pages 10687--10698, 2020.

\bibitem{Cubu18}
Ekin~D Cubuk, Barret Zoph, Dandelion Mane, Vijay Vasudevan, and Quoc~V Le.
\newblock Autoaugment: Learning augmentation policies from data.
\newblock {\em arXiv preprint arXiv:1805.09501}, 2018.

\bibitem{Cubu20}
Ekin~D Cubuk, Barret Zoph, Jonathon Shlens, and Quoc~V Le.
\newblock Randaugment: Practical automated data augmentation with a reduced
  search space.
\newblock In {\em Proceedings of the IEEE/CVF conference on computer vision and
  pattern recognition workshops}, pages 702--703, 2020.

\bibitem{Mull21}
Samuel~G M{\"u}ller and Frank Hutter.
\newblock Trivialaugment: Tuning-free yet state-of-the-art data augmentation.
\newblock In {\em Proceedings of the IEEE/CVF International Conference on
  Computer Vision}, pages 774--782, 2021.

\bibitem{Wang22}
He~Wang, Hongyi Sun, Yi~Fang, Shuo Li, Ming Feng, and Renzhi Wang.
\newblock A workflow for computer-aided diagnosis of glaucoma.
\newblock In {\em 2022 IEEE International Symposium on Biomedical Imaging
  Challenges (ISBIC)}, pages 1--4. IEEE, 2022.

\bibitem{Khad22}
Firas Khader, Christoph Haarburger, J{\"o}rg-Christian Kirr, Marcel Menke,
  Jakob~Nikolas Kather, Johannes Stegmaier, Christiane Kuhl, Sven Nebelung, and
  Daniel Truhn.
\newblock Elevating fundoscopic evaluation to expert level-automatic glaucoma
  detection using data from the airogs challenge.
\newblock In {\em 2022 IEEE International Symposium on Biomedical Imaging
  Challenges (ISBIC)}, pages 1--4. IEEE, 2022.

\bibitem{Arau22}
Teresa Ara{\'u}jo, Guilherme Aresta, and Hrvoje Bogunovi{\'c}.
\newblock Deep dirichlet uncertainty for unsupervised out-of-distribution
  detection of eye fundus photographs in glaucoma screening.
\newblock In {\em 2022 IEEE International Symposium on Biomedical Imaging
  Challenges (ISBIC)}, pages 1--5. IEEE, 2022.

\bibitem{Ziao18}
Tete Xiao, Yingcheng Liu, Bolei Zhou, Yuning Jiang, and Jian Sun.
\newblock Unified perceptual parsing for scene understanding.
\newblock In {\em Proceedings of the European conference on computer vision
  (ECCV)}, pages 418--434, 2018.

\bibitem{Shar21}
Gilad Sharir, Asaf Noy, and Lihi Zelnik-Manor.
\newblock An image is worth 16x16 words, what is a video worth?
\newblock {\em arXiv preprint arXiv:2103.13915}, 2021.

\bibitem{Liu21}
Ze~Liu, Yutong Lin, Yue Cao, Han Hu, Yixuan Wei, Zheng Zhang, Stephen Lin, and
  Baining Guo.
\newblock Swin transformer: Hierarchical vision transformer using shifted
  windows.
\newblock In {\em Proceedings of the IEEE/CVF International Conference on
  Computer Vision}, pages 10012--10022, 2021.

\bibitem{Touv21}
Hugo Touvron, Matthieu Cord, Matthijs Douze, Francisco Massa, Alexandre
  Sablayrolles, and Herv{\'e} J{\'e}gou.
\newblock Training data-efficient image transformers \& distillation through
  attention.
\newblock In {\em International Conference on Machine Learning}, pages
  10347--10357. PMLR, 2021.

\bibitem{Tan19}
Mingxing Tan and Quoc Le.
\newblock Efficientnet: Rethinking model scaling for convolutional neural
  networks.
\newblock In {\em International conference on machine learning}, pages
  6105--6114. PMLR, 2019.

\bibitem{Tan21}
Mingxing Tan and Quoc Le.
\newblock Efficientnetv2: Smaller models and faster training.
\newblock In {\em International conference on machine learning}, pages
  10096--10106. PMLR, 2021.

\bibitem{Joch20}
Glenn Jocher, Alex Stoken, Jirka Borovec, Liu Changyu, Adam Hogan, L~Diaconu,
  F~Ingham, J~Poznanski, J~Fang, L~Yu, et~al.
\newblock ultralytics/yolov5.
\newblock {\em Zenodo}, 2020.

\bibitem{Aimy22}
Temirgali Aimyshev and Kanat Eleusiz.
\newblock Glaucoma detection algorithm for the artificial intelligence for
  robust glaucoma screening challenge, 2022.

\bibitem{Doso20}
Alexey Dosovitskiy, Lucas Beyer, Alexander Kolesnikov, Dirk Weissenborn,
  Xiaohua Zhai, Thomas Unterthiner, Mostafa Dehghani, Matthias Minderer, Georg
  Heigold, Sylvain Gelly, et~al.
\newblock An image is worth 16x16 words: Transformers for image recognition at
  scale.
\newblock {\em arXiv preprint arXiv:2010.11929}, 2020.

\bibitem{Le22}
Tien-Dung Le.
\newblock Combination of supervised learning and unsupervised learning to
  detect ungradable images in the airogs challenge, 2022.

\bibitem{Huan17}
Gao Huang, Zhuang Liu, Laurens Van Der~Maaten, and Kilian~Q Weinberger.
\newblock Densely connected convolutional networks.
\newblock In {\em Proceedings of the IEEE conference on computer vision and
  pattern recognition}, pages 4700--4708, 2017.

\bibitem{Gald22}
Adrian Galdran, Gustavo Carneiro, and Miguel A.~Gonzlez Ballester.
\newblock Open-set glaucoma screening from eye fundus images: Domain knowledge
  to the rescue, 2022.

\bibitem{Howa17}
Andrew~G Howard, Menglong Zhu, Bo~Chen, Dmitry Kalenichenko, Weijun Wang,
  Tobias Weyand, Marco Andreetto, and Hartwig Adam.
\newblock Mobilenets: Efficient convolutional neural networks for mobile vision
  applications.
\newblock {\em arXiv preprint arXiv:1704.04861}, 2017.

\bibitem{Puth22}
Densen Puthussery, P~S Hrishikesh, R~G Devika, and C~V Jiji.
\newblock A self-supervised approach for glaucoma screening, 2022.

\bibitem{Lin17}
Tsung-Yi Lin, Priya Goyal, Ross Girshick, Kaiming He, and Piotr Doll{\'a}r.
\newblock Focal loss for dense object detection.
\newblock In {\em Proceedings of the IEEE international conference on computer
  vision}, pages 2980--2988, 2017.

\bibitem{Oza18}
Poojan Oza and Vishal~M Patel.
\newblock One-class convolutional neural network.
\newblock {\em IEEE Signal Processing Letters}, 26(2):277--281, 2018.

\bibitem{Yang22}
Zekang Yang, Hong Liu, and Zihao Shang.
\newblock Deep learning for referable glaucoma screening and
  out-of-distribution detection, 2022.

\bibitem{Kond22}
Satoshi Kondo, Satoshi Kasai, and Kosuke Hirasawa.
\newblock Computer aided diagnosis and out-of-distribution detection in
  glaucoma screening using color fundus photography.
\newblock {\em arXiv preprint arXiv:2202.11944}, 2022.

\bibitem{Bell21}
Irwan Bello, William Fedus, Xianzhi Du, Ekin~Dogus Cubuk, Aravind Srinivas,
  Tsung-Yi Lin, Jonathon Shlens, and Barret Zoph.
\newblock Revisiting resnets: Improved training and scaling strategies.
\newblock {\em Advances in Neural Information Processing Systems},
  34:22614--22627, 2021.

\bibitem{Liu20}
Weitang Liu, Xiaoyun Wang, John Owens, and Yixuan Li.
\newblock Energy-based out-of-distribution detection.
\newblock {\em Advances in Neural Information Processing Systems},
  33:21464--21475, 2020.

\bibitem{Sun21}
Yiyou Sun, Chuan Guo, and Yixuan Li.
\newblock React: Out-of-distribution detection with rectified activations.
\newblock {\em Advances in Neural Information Processing Systems}, 34:144--157,
  2021.

\bibitem{Wang22a}
Edward Wang, Ashritha Durvasula, Daisy Deng, Asaanth Sivajohan, Edward Ho, and
  Kevin Lane.
\newblock Ensemble network for glaucoma screening in airogs challenge, 2022.

\bibitem{Hu18}
Jie Hu, Li~Shen, and Gang Sun.
\newblock Squeeze-and-excitation networks.
\newblock In {\em Proceedings of the IEEE Conference on Computer Vision and
  Pattern Recognition}, pages 7132--7141, 2018.

\bibitem{Simo14}
Karen Simonyan and Andrew Zisserman.
\newblock Very deep convolutional networks for large-scale image recognition.
\newblock {\em arXiv preprint arXiv:1409.1556}, 2014.

\bibitem{Szeg16}
Christian Szegedy, Vincent Vanhoucke, Sergey Ioffe, Jon Shlens, and Zbigniew
  Wojna.
\newblock Rethinking the inception architecture for computer vision.
\newblock In {\em Proceedings of the IEEE conference on computer vision and
  pattern recognition}, pages 2818--2826, 2016.

\bibitem{King13}
Diederik~P Kingma and Max Welling.
\newblock Auto-encoding variational bayes.
\newblock {\em arXiv preprint arXiv:1312.6114}, 2013.

\bibitem{Hera22}
J\'{o}nathan Heras, Didac Royo, and Miguel~\'{A}ngel Zapata.
\newblock A good closed-set classifier is all you need for the airogs
  challenge, 2022.

\bibitem{Vaze21}
Sagar Vaze, Kai Han, Andrea Vedaldi, and Andrew Zisserman.
\newblock Open-set recognition: A good closed-set classifier is all you need.
\newblock {\em arXiv preprint arXiv:2110.06207}, 2021.

\bibitem{Demp68}
Arthur~P Dempster.
\newblock A generalization of bayesian inference.
\newblock {\em Journal of the Royal Statistical Society: Series B
  (Methodological)}, 30(2):205--232, 1968.

\bibitem{Selv17}
Ramprasaath~R Selvaraju, Michael Cogswell, Abhishek Das, Ramakrishna Vedantam,
  Devi Parikh, and Dhruv Batra.
\newblock Grad-cam: Visual explanations from deep networks via gradient-based
  localization.
\newblock In {\em Proceedings of the IEEE International Conference on Computer
  Vision}, pages 618--626, 2017.

\bibitem{Arik22}
Mustafa Arikan.
\newblock Multi-model ensemble for robust glaucoma screening, 2022.

\bibitem{Szeg17}
Christian Szegedy, Sergey Ioffe, Vincent Vanhoucke, and Alexander~A Alemi.
\newblock {Inception-v4, Inception-ResNet and the impact of residual
  connections on learning}.
\newblock In {\em Thirty-first AAAI Conference on Artificial Intelligence},
  2017.

\bibitem{Chan22}
Yeong~Chan Lee, Hyun~Bin Cho, and Yoon~Ho Choi.
\newblock Classification for referable glaucoma with fundus photographs using
  multimodal deep learning, 2022.

\bibitem{Ronn15}
Olaf Ronneberger, Philipp Fischer, and Thomas Brox.
\newblock {U-Net}: Convolutional networks for biomedical image segmentation.
\newblock In {\em International Conference on Medical image computing and
  computer-assisted intervention}, pages 234--241. Springer, 2015.

\bibitem{Gal16}
Yarin Gal and Zoubin Ghahramani.
\newblock Dropout as a bayesian approximation: Representing model uncertainty
  in deep learning.
\newblock In {\em international conference on machine learning}, pages
  1050--1059. PMLR, 2016.

\bibitem{Qayy22}
Abdul Qayyum, Moona Mazher, and Imran Razzak.
\newblock Convnexts and vision transformer based framework for glaucoma
  screening, 2022.

\bibitem{Liu22}
Zhuang Liu, Hanzi Mao, Chao-Yuan Wu, Christoph Feichtenhofer, Trevor Darrell,
  and Saining Xie.
\newblock A convnet for the 2020s.
\newblock In {\em Proceedings of the IEEE/CVF Conference on Computer Vision and
  Pattern Recognition}, pages 11976--11986, 2022.

\bibitem{Biba21}
Koby Bibas, Meir Feder, and Tal Hassner.
\newblock Single layer predictive normalized maximum likelihood for
  out-of-distribution detection.
\newblock {\em Advances in Neural Information Processing Systems},
  34:1179--1191, 2021.

\bibitem{Lemi22}
Hans~G Lemij, Coen de~Vente, Clara S{\'a}nchez, Jorge Cuadros, Nicolas Jaccard,
  and Koen Vermeer.
\newblock Glaucomatous features in fundus photographs of eyes with 'referable
  glaucoma' of a large population based labeled data set for training an
  artificial intelligence ({AI}) algorithm for glaucoma screening.
\newblock In {\em Association for Research in Vision and Ophthalmology},
  volume~63, pages 2041--A0482. The Association for Research in Vision and
  Ophthalmology, 2022.

\end{thebibliography}

\end{document}